\pdfoutput=1 
\documentclass[sort&compress,times,preprint]{elsarticle}
\usepackage[margin=20mm]{geometry}
\usepackage[ascii]{inputenc}
\usepackage[english]{babel}
\usepackage{amsmath}
\usepackage{mathtools, cuted}
\usepackage{graphicx}
\usepackage{subcaption}

\makeatletter
\def\ps@pprintTitle{%
   \let\@oddhead\@empty
   \let\@evenhead\@empty
   \let\@oddfoot\@empty
   \let\@evenfoot\@oddfoot
}
\makeatother

\DeclareMathOperator{\diag}{diag}

\begin{document}

\begin{frontmatter}

  \title{
  The lifetime of micron scale topological chiral magnetic states with atomic resolution
  }
   
  \author[spbu]{I S Lobanov\corref{cor}\fnref{spbu}\fnref{phystech}}
  \cortext[cor]{Corresponding author}
  \ead{lobanov.igor@gmail.com}

  \author[spbu]{V M Uzdin\fnref{spbu}\fnref{phystech}}
  
  \address[spbu]{Department of Physics, St. Petersburg State University, St. Petersburg 198504, Russia}
  \address[phystech]{Faculty of Physics, ITMO University, St. Petersburg 197101, Russia}

\begin{abstract}
A new method for the numerical computation of the lifetimes of magnetic states within harmonic transition state theory (HTST) has been developed. 
In the simplest case, the system is described by a Heisenberg-like Hamiltonian with short- range interaction. Calculations are performed in Cartesian coordinates. 
Constraints on the values of magnetic moments are taken into account using Lagrange multipliers. 
The pre-exponential factor in the Arrhenius law in HTST is written in terms of the determinants of the Hessian of energy 
at the minima and saddle points on the multidimensional energy surface. 
An algorithm for calculating these determinants without searching for eigenvalues of the Hessian but using recursive relations is proposed. 
The method allows calculating  determinants for  systems containing millions of magnetic moments. 
This makes it possible to calculate the pre-exponential factor and estimate the lifetimes of micron-scale topological structures with atomic resolution, 
which until now has been impossible using standard approaches. The accuracy of the method is demonstrated by calculating 2D and 3D skyrmionic structures.
\end{abstract}


\end{frontmatter}

\section{Introduction}

The thermal stability of magnetic states is a key problem for the use of nano- and
microstructures in the new generation of magnetic memory, in logical and neuromorphic devices. 
One way to create stable magnetic nanostructures involves the utilization of artificial topological systems \cite{ref1,ref2}. 
For these systems, one can determine the topological charge, which is an integer that remains constant with a continuous change in magnetization. 
This should protect the system with continuous magnetization from thermal fluctuations and random disturbances \cite{ref3,ref4,ref5}. 

For magnetic moments localized at the sites of the discrete lattice, topological arguments, strictly speaking, are absent. 
Stability estimation and lifetime calculation for such magnetic structures are a complex problems that have not yet been resolved for most states of interest for applications. 

For this purpose, in principle, one can use the transition state theory (TST) for magnetic degrees of freedom \cite {ref6}, as well as the Kramers-Langer theory \cite {ref7, ref8}. 
These approaches are based on the analysis of the multidimensional energy surface of the system, the search for minima on it corresponding to locally stable states, the construction of minimum energy paths (MEP)
between minima, which determine the most probable transition scenarios \cite{ref9,ref10, ref11,ref12,ref13}. 
The state with maximum energy along the MEP is a saddle point on the energy surface. 
The difference in energy at the saddle point and the minimum gives the activation energy for the corresponding magnetic transition \cite{ref14,ref15}. 
An analytical expression for the rate of magnetic transitions corresponding to the Arrhenius law can be derived 
in the harmonic approximation, when the shapes of the energy surface at the minima and saddle points are approximated by a quadratic
polynomial in all variables \cite{ref6,ref11}.
The expression for the pre-exponential factor in this law contains the product of the eigenvalues of the Hessian of energy at the minimum and
saddle point \cite{ref6,ref17}. 

This approach was implemented to calculate the lifetimes of skyrmions in the fcc-Pd / Fe / Ir(111)
system \cite{ref18, ref19, ref20}, where the skyrmion (Sk) size was several nanometers,
and the dimension of the energy surface did not exceed several thousand. 
However, such Sks are stable only at low temperatures of few tens of Kelvin.
Sks, which are stable at room temperature, are much larger, and often they are no longer two-dimensional, 
but three-dimensional structures \cite{ref21,ref22}.
Even in antiferro- and ferrimagnetic materials in which 
Sks of small radius were detected at room temperature \citep{ref23,ref24}, the total number of magnetic 
moments that make up the structure is of order tens of thousands or more.
In ferromagnets, Sks can contain millions of magnetic moments.
The number of degrees of freedom and the dimension of the energy surface in this case may exceeds several millions, 
and the task of calculating the activation energy and the pre-exponential factor is qualitatively more complex. 

An approach for MEP computation for a Sk having a several millions-di\-men\-sion\-al energy manifold was proposed using 
truncated minimum energy path method \cite{ref25}. 
It takes into account the fact that the noncollinear structure in transition state has a much smaller size than in the ground state. 
But the calculations of the pre-exponential factor for such systems have not yet been carried out. 
This is due to the difficulties of finding eigenvalues 
of a matrix with a rank of several million with the necessary accuracy in a commonly used expression for pre-exponential factor. 
In this article, we propose a method to compute the pre-exponential factor in the Arrhenius law 
that does not require knowledge of the complete set of Hessian eigenvalues, 
but utilizes a recursion relation for the Hessian determinant
reducing the problem to computation of determinants of much smaller matrices.

The computation of determinants of large matrices is a challenging task appearing as a part of
analysis in many physical problems. 
For general matrix of size $N\times N$ complexity of determinant computation is $O(N^3)$ \cite{PYS97}. 
Therefore typically only systems with relatively small number of degrees of freedom, about $10^4\-- 10^5$, 
can be numerically analyzed \cite{ref11,ref12,ref13,ref15,ref17,ref18,ref19}. 
However, if the matrix has special structure, the complexity can be reduced,
in particular if the matrix is banded with a small number $K$ of diagonals, the determinant calculation time is $O(NK^2)$ \cite{GvL89}.
This makes possible to analyze systems with millions of degrees of freedom. 
Such situation occurs when calculating the lifetime of topological magnetic systems, if only short-range
interactions between the magnetic moments are taken into account. 
The Hessian matrix block structure in the case is known in advance, 
since it is determined by pairs of directly interacting magnetic moments. 
Therefore there is no need to store the entire matrix, but only nonzero elements have to be calculated and saved. 
Corresponding algorithms for calculating determinants for the Hessian of energy are
developed in this article.

Exploiting block-diagonal structure of matrix is quite known approach in many applied fields, 
but seems to be new in the context of lifetime estimation of magnetic systems.
Block LU diagonalization is described in details including error analysis in books 
\cite{H02}.
Explicit formulas for block tridiagonal matrices 
are given by Molinary \cite{M08}.
The approach of Molinary can be interpreted as 
an discrete version of technique used in \cite{C85}
 for computation of functional determinants.
However, this approach requires inversion of the off-diagonal elements of the Hessian matrix, which leads to huge round-off errors, 
that makes the method inapplicable for the quantitative description of magnetic systems.

This article focuses on the development of a fast and robust algorithm for calculation of the lifetime of magnetic states. 
An important aspect of this problem is the choice of coordinates minimizing error of representation of magnetic moments,
suitable for high-performance computations. 
Since the length of the magnetic moments is constant in the Heisenberg-like model, the natural choice is to use
two spherical angles to define the orientation of the moments. However spherical coordinates suffer from a loss of accuracy near the poles. Several other options have been proposed to overcome problem, e.g. use of stereographic projections (ATLAS method) \cite{RBBK15}, orthogonal matrix transformation \cite{IDTUJ20}, etc.

In the article we show how to compute determinants
using Cartesian coordinates, the same coordinates
that used in the definition of energy of the system.
Combining the proposed method of determinant
evaluation with fast method of dynamic prefactor
computation suggested in \cite{ref28},
a straightforward and fast method of lifetime computation is obtained.

The article is organized as follows.
In the second section, we introduce the expression for the energy of the magnetic system in the framework of the generalized Heisenberg model in harmonic approximation, 
and derive an explicit formula for the Hessian matrix for the energy. 
The third section presents a modified formula for the magnetic transition rate within the HTST.  In the harmonic approximation, the pre-exponential factor is more convenient to express through Hessian determinants instead of its eigenvalues. 
The dynamic factor can also be rewritten in a form independent of the local basis chosen at the saddle point of the energy surface, 
which is more convenient for numerical calculations. 
Then we give a method for calculating the determinants of the Hessian block matrices with a fixed relatively small number of nonzero diagonals. 
We first discuss the simpler case of free boundary conditions, and then we proceed to the boundary conditions of a periodic type. 
We develop formulas for both cases, and also show how to use the Woodbury-Morison-Sherman equation to reduce a periodic case to a free one. 
Finally, we give a benchmark computing pre-exponential factor for decay of magnetic Sks, comparing the result with commonly used approaches.

\section{Energy of magnetic system. Gradient and Hessian of energy}

We will use generalized Heisenberg model, which describes magnetic moments as  classical vectors $\mathbf M_n=\mu_n\mathbf S_n$, 
localized on sites of a crystal lattice.
The values of magnetic moments $\mu_n$ are supposed to be constant, and magnetic configuration is determined by set of unit vectors $\mathbf S_n$.
The energy $E$ of the system, depending on the configuration $\mathbf S=(\mathbf S_n)$, 
$n=1\ldots N$, is written as

\begin{equation}\label{eq1}
E = E[\mathbf S] = E_{ex} + E_{DM} + E_a + E_Z
\end{equation}

Here Heisenberg term $E_{ex}$ and energy of Dzyaloshinskii-Morya interaction (DMI) $E_{DM}$ 
are quadratic forms in variables $\bf{S_n}$:

\begin{equation}\label{eq2}
E_{ex}=-\sum_{\langle n,m\rangle}J_{nm}(\mathbf {S}_n\cdot \mathbf {S}_m),\quad 
E_{DM}=-\sum_{\langle n,m\rangle}\mathbf{D}_{nm}\cdot[\mathbf {S}_n\times \mathbf {S}_m]
\end{equation}
the sums are taken over all pairs $\langle n,m \rangle$ of interacting moments. 
Often only pairs of nearest neighbor moments are considered, since these interactions decrease with distance. 
However, in ref. \cite{ref26} it is argues that taking into account exchange frustrations and 
the inclusion of several exchange parameters $J_{nm}$
can lead to a significant increase in the lifetime of isolated Sks according to their calculations. 
Also several different DMI-vectors $D_{nm}$ can be considered \cite{ref27}. 

The anisotropy energy $E_a$ contains quadratic terms associated with the local on-site magnetic moment $\mathbf M_n$:

\begin{equation}\label{eq3}
E_a=-\sum _n K  \mathbf{(\mathbf e_K\cdot \mathbf S_n)}^2
\end{equation}
where $K$ is anisotropy constant, and unit vector $\mathbf e_K$ is along the anisotropy axis. 
There may be several anisotropies, in which case their energies are added up. 
Multiple anisotropies do not present any difficulties in our study, 
so we will discuss a single easy axis anisotropy $K>0$ for  shortness. 

Zeeman energy $E_Z$ is an interaction with an external local magnetic field $\mathbf B_n$,  
it is linear with respect to $\mathbf M_n$:

\begin{equation}\label{eq4}
E_Z=-\sum_n\mu_n\mathbf B_n\cdot\mathbf S_n
\end{equation}
Quadratic terms (\ref{eq2}) and (\ref{eq3}) can be written in a matrix form, which is more suitable for
further analysis: 

\begin{equation}\label{eq5}
E_{ex}+E_{DM} = \frac12\sum _{n,m}\mathbf S_n\cdot \widehat A_{nm}\mathbf S_m.
\end{equation}

\begin{equation*}
E_a=\frac12\sum_n\mathbf S_n\cdot \widehat  C_n\mathbf S_n
\end{equation*}
where the matrices $\widehat A_{nm}$ and $\widehat C_{n}$ have the following explicit forms:

\begin{equation}\label{eq6}
\widehat A_{nm} = -J_{nm}I_3-[\mathbf D_{nm}]_{\times},\quad
\widehat C_n=-2K|\mathbf e_K\rangle\langle\mathbf e_K|
\end{equation}
Here $[\mathbf a]_\times$ is the skew-Hermitian matrix corresponding to the vector $\mathbf a$:
\[
  [\mathbf a]_{\times}=\begin{pmatrix}0&-a_z&a_y\\a_z&0&-a_x\\-a_y&a_x&0\end{pmatrix},
\]
and  
\[
  |\mathbf a\rangle\langle \mathbf a|
  =\begin{pmatrix}a_x^2&a_xa_y&a_xa_z\\a_xa_y&a_y^2&a_ya_z\\a_xa_z&a_ya_z&a_z^2\end{pmatrix}.
\]

In what followed we do not need an explicit form of the matrices; therefore, other types of interactions
containing only quadratic terms can be included in the scheme. 
Below we use that the matrices (\ref{eq6}) have the following properties:  
$\widehat A_{nm}=\widehat  A_{mn}^T$, $\widehat  C_n=\widehat  C_n^T$ for all $n$, $m$. 
If only nearest-neighbours are taken into account, the Hessian matrix has block-diagonal form.
The method developed in Section \ref{sec:detlu} is also applicable for short-range interactions with 
a finite number of interacting spins (even beyond nearest-neighbours), which result in block-banded structure of the matrix. 

The dynamics of magnetic states and their entropy are determined by gradient and Hessian of energy.
In the absence of damping, the equation of motion for the magnetic system is given by the Landau-Lifshitz equation
\[
  \frac{d \mathbf S_n}{dt} = -\frac{\gamma}{\mu_n} \mathbf S_n\times \mathbf H^{eff}_n[\mathbf S], 
\]
Here $\gamma$ is gyromagnetic ratio. Effective field $\mathbf H_{eff}$,
is expressed in terms of gradient of energy:
\[
  \mathbf H^{eff}_n[\mathbf S]=-\frac{\partial E[\mathbf S]}{\partial \mathbf S_n}
  =-\nabla_n E[\mathbf S].
\]
Vector of gradient $\nabla_n E[\mathbf S]$ not necessary belong
to the tangent space to the sphere $\mathbf S_n^2=1$,
but only projection of the gradient 
to the tangent space $P\nabla E[\mathbf S]$ affects the motion:
\[
  P=\diag_n P_n,\quad 
  P_n \mathbf x = \mathbf x-\mathbf S_n(\mathbf S_n\cdot \mathbf x).
\]
The gradient also points in the direction of the fastest increase
of energy, making it useful for optimization
problems, in particular for computation of metastable states 
and minimum energy paths. 
When a magnetic state moves along the energy gradient $\nabla_{\mathbf S}E$, the constraints imposed on the system, such as the constancy of the values of  individual magnetic moments, may not be met.
This problem does not arise if we use coordinates on the manifold, defined by the constraints conditions. 
However, the introduction of global coordinates without singularities on the manifold is impossible and local coordinates 
have to be used to obtain good accuracy.
Here we propose a method for introducing local coordinates near the state $\mathbf S^0$ in a natural way.

For a particular state $\mathbf S^0 $, the tangent space to the manifold, which is determined by restrictions on the value of the magnetic moment at the point $\mathbf S^0$, is formed by all vectors
$ \mathbf \Delta \in \mathbb R^{3N}$, satisfying the constraints
$ \mathbf \Delta_n \cdot \mathbf S^0_n = 0$.

The tangent space coordinates $\Delta$ can be transformed 
to the coordinates in a neighborhood of $\mathbf S^0$ 
on the  manifold as follows:

\begin{equation*}
\mathbf S_n(\mathbf \Delta)=\frac{\mathbf S_n^0+\mathbf \Delta_n}{\|\mathbf S_n^0+\mathbf \Delta_n\|}
\end{equation*}
The nonlinear  mapping is quite complex for analysis,
however it can be simplified if only few terms of 
series expansion for small $\mathbf \Delta$ is required:
\begin{equation}\label{eq7}  
\mathbf S_n=\frac{\mathbf S_n^0+\mathbf \Delta_n}{\sqrt{1+(\mathbf \Delta_n)^2}}
=\mathbf S_n^0+\mathbf \Delta_n-\frac{\mathbf \Delta_n^2}2 \mathbf S_n^0+o(\mathbf \Delta_n)^2
\end{equation}
First two terms on the right-hand side defines the tangent space, 
and the third term takes into account curvature of the manifold.

Convenient $2N$-dimensional coordinates $\xi$ can be introduced
fixing arbitrary orthonormal basis $\mathbf e$ in the tangent spaces:
\begin{equation}\label{eq:tbasis}
  \mathbf \Delta_n=\xi_n^1 \mathbf e_n^1+\xi_n^2 \mathbf e_n^2,\quad
  \mathbf e^1_n\cdot \mathbf S_n^0=\mathbf e^2_n\cdot \mathbf S_n^0
  =\mathbf e^1_n\cdot \mathbf e^2_n=0,\quad
  (\mathbf e^k_n)^2=1.
\end{equation}
We denote by $\Pi[\mathbf S^0]$ the embedding $\Pi [\mathbf S^0]\mathbf \xi=\mathbf \Delta$ 
of the tangent space in $\mathbb R^{3N}$.

Using coordinates $\mathbf \Delta$, it is easy to compute 
variation of energy on the manifold.
It is worth noting that though $\mathbf\Delta$ 
belongs to the same space $\mathbb R^{3N}$ as $\mathbf S$,
they have different constraints and 
energy has different expressions in their terms.

The energy (\ref{eq1}) introduced above can be written as a quadratic form 

\begin{equation}\label{eq:energy_gen}
  E[\mathbf S] = \frac12 \mathbf S\cdot \mathcal A \mathbf S-\mu \mathbf B\cdot \mathbf S,
\end{equation}
where 
$\mathcal A_{nm}=\widehat A_{nm}, n \ne m $
and 
$\mathcal A_{nn}=\widehat C_{n}$.
Since matrix $\mathcal A$ is Hermitian, 
gradient of energy is $\nabla E[\mathbf S]=\mathcal A \mathbf S-\mu \mathbf B$.
Using the decomposition (\ref{eq7}), we can 
express energy in harmonic approximation.
To do this we write gradient in terms of $\mathbf \Delta$:
\[
  \nabla E[\mathbf S] = \nabla E[\mathbf S^0] 
    + \mathcal A\mathbf \Delta 
    - \frac12 \mathcal A D\mathbf S^0 +o(\mathbf \Delta^2)\text{ as }\mathbf \Delta\to0,
\]
where $\nabla E[\mathbf S^0]=\mathcal A \mathbf S^0-\mu \mathbf B$, 
$D=\diag_n \mathbf \Delta_n^2$.
Expression for energy (\ref{eq:energy_gen}) now can be rewritten as
\[
  E[\mathbf S] = \frac12 \mathbf S\cdot (\nabla E[\mathbf S]-\mu \mathbf B),
\]
and we get an expansion of energy with terms of the first and second order in $\Delta$:
\[
  E[\mathbf S] = E[\mathbf S^0] + E_1 + E_2 + o(\Delta^2),
\]
where
\[
  E_1 = \frac12 \mathbf \Delta\cdot(\nabla E[\mathbf S^0]-\mu \mathbf B)
  +\frac12\mathbf S^0\cdot\mathcal A\mathbf \Delta,
\]
\[
  E_2 = -\frac14\mathbf S^0\cdot\mathcal A D\mathbf \Delta
  -\frac14\mathbf \Delta\cdot\mathcal AD\mathbf S^0 
  -\frac14(D\mathbf S^0)\cdot\mathcal A\mathbf \Delta
  -\frac14(D\mathbf \Delta)\cdot(\nabla E[\mathbf S^0]-\mu \mathbf B). 
\]
Using definition of the gradient and that $\mathcal A$ 
and $D$ are Hermitian, the expressions are simplified to:
\[
  E_1 = \mathbf \Delta\cdot(\mathcal A\mathbf S^0-\mu \mathbf B),
\]
\[
  E_2 = \frac12\mathbf \Delta\cdot\mathcal A\mathbf \Delta
  -\frac12(\nabla E[\mathbf S^0])\cdot D\mathbf S^0.
\]
The last term in $E_2$ is a quadratic form in $\mathbf \Delta$.
Indeed, if we introduce the Lagrange multipliers 
$\lambda_n=\mathbf S^0_n\cdot\nabla_n E[\mathbf S^0]$ 
and define the diagonal matrix  $\Lambda=\diag_n \lambda_n$,
it can be written as:
\[
  (\nabla E[\mathbf S^0])\cdot D\mathbf S^0=\mathbf \Delta\cdot \Lambda\mathbf \Delta.
\]
Collecting all terms, we obtain harmonic approximation for energy:
\[
  E[\mathbf S] 
  = E[\mathbf S^0] + \mathbf \Delta\cdot\nabla E[\mathbf S^0]
  + \frac12\mathbf \Delta \cdot(\mathcal A-\Lambda[\mathbf S^0])\mathbf \Delta
  + o(\mathbf \Delta^2).
\]
The matrix of quadratic term of energy 
is Hessian matrix $\widehat H$:
\begin{equation}\label{eq:AminusLambda}
  \widehat H[\mathbf S^0]=\mathcal A-\Lambda[\mathbf S^0],\quad
  \widehat H_{nn} = \widehat C_n-\lambda_n,\quad
  \widehat H_{nm} = \widehat A_{nm}.
\end{equation}

Since energy $E$ is a quadratic form of variables $\mathbf S$,
its Hessian with respect to $\mathbf S$ is simply the matrix $\mathcal A$.
To take into account constraints we introduced coordinates 
$\mathbf \Delta$ in constraints manifold, and it turns out that
Hessian of energy in coordinates $\mathbf \Delta$ coincide 
with the matrix $\mathcal A$ minus a diagonal matrix 
of Lagrange multipliers, see (\ref{eq:AminusLambda}),
which makes it easy to compute the Hessian.  
 However we should remember that vectors $\mathbf \Delta_n$
must be orthogonal to $\mathbf S^0_n$ in computations above.
The conditions are easy to satisfy doing energy minimization (computation of metastable states),
hence the Hessian matrix can be readily used to implement 
e.g. non-linear conjugate gradient method.
However, we can get rid of the constraints restricting
the matrix to $2N$ space resulting in 
the final Hessian matrix using the embedding $\Pi[\mathbf S^0]$ defined next to the equation (\ref{eq:tbasis}):

\begin{equation}\label{eq:projhess}
\mathcal H[\mathbf S^0]=\Pi^T[\mathbf S^0]\; \widehat H[\mathbf S^0]\; \Pi[\mathbf S^0].
\end{equation}

\section{Rates of magnetic transitions in HTST}
\label{sec:rate}

TST for magnetic degrees of freedom presupposes that magnetic moments are classical objects
with energy defined, for example, by the equation (\ref{eq1}), and that magnetic transitions are slow with respect to magnetic vibrations so
that the Boltzmann distribution is established and maintained  in the whole available  region in the energy surface up to and including the transition state.
Moreover in TST the trajectories on the energy surface suppose to be reactive, 
i.e. go from initial to final states crossing the dividing surface only once \cite{ref10}.

Initially the system is in the local equilibrium in a neighborhood of the minimum on the energy surface $\mathbf S^m$. 
In accordance with the Boltzmann distribution the probability $dP$
of finding the system in an element of the configuration space $d\mathbf S$
with energy $E$ can be written as

\begin{equation*}
dP=Z^{-1}\exp\left(-\frac{E[\mathbf S]}{k_BT}\right){d\mathbf S}
\end{equation*}
The normalization factor $Z$ can be found in the harmonic approximation, 
if we assume that the energy surface at the minimum is well described by a quadratic dependence on all variables. 
In a neighborhood of the minimum $\mathbf S^m$,
where $\mathbf \nabla E[\mathbf S^m]=0$, 
the energy surface is locally described by the Hessian matrix 
$\widehat H$ of energy at $\mathbf S^m$:
\[
E[\mathbf S]=E[\mathbf S^m]+\frac12 \mathbf \Delta\cdot \mathcal H[\mathbf S^m]\mathbf \Delta+o(\mathbf \Delta^2),
\]
where an explicit form of Hessian is derived above.  
Denote by $\mathbf e_m^{(i)}$ an orthonormal family of eigenvectors of 
$\mathcal H$ in configuration $\mathbf S^m$:
$$\mathcal H[\mathbf S^m] \mathbf e_m^{(i)}=\zeta _m^{(i)} \mathbf e_m^{(i)},\quad
\mathbf e_m^{(i)}\cdot \mathbf e_m^{(j)}=\delta_{ij}.$$
Expanding the increment $\mathbf \Delta$ over the basis 
$\mathbf e_m^{(i)}\colon\mathbf \Delta=\sum _i\mathbf \Delta_m^{(i)} \mathbf e_m^{(i)}$ 
we obtain the following expansion for the energy:

\begin{equation}\label{eq13}
E[\mathbf S]=E[\mathbf S^m]+\sum_i\zeta_m^{(i)}(\mathbf \Delta_m^{(i)})^2+o(\Delta^2),
\end{equation}
At a local minimum on the energy surface, all eigenvalues are positive: $\zeta_m^{(i)}\geq0$. 
If they are large enough so that the probability of finding the system outside the vicinity of $\mathbf S^m$
is negligible, the normalization factor $Z$ can be calculated analytically:

\begin{equation}\label{eq14}
e^{\frac{E[\mathbf S^m]}{k_BT}}Z=\int_{\mathbb R^{2N}}
\exp\left(-\frac{\sum_i\zeta_m^{(i)}(\mathbf \Delta_m^{(i)})^2}{k_BT}\right)d\mathbf \Delta
=\frac{(\pi k_BT)^N}{\sqrt{\det\widehat H[\mathbf S^m]}},
\end{equation}

Within TST magnetic transition go through the transition state (TS) 
which is the first order saddle point $\mathbf S^{TS}$ on the energy surface. 
Hessian $\mathcal H[\mathbf S^{TS}]$ here has exactly one negative eigenvalue. 
We will not consider in the article so-called zero modes (cyclic variables), 
the variables of which the energy is independent. 
Such modes should be considered separately \cite{ref18,ref19}.
For convenience we will enumerate eigenvalues in the increasing order: 
$\zeta _{TS}^{(1)}<0<\zeta _{TS}^{(2)}\leq\cdots\leq\zeta_{TS}^{(2N)}$. 
Rate of magnetic transitions $\kappa$, which is inverse lifetime of magnetic state $1/\tau$, 
can be estimated as product of probability being at transition state near
the dividing surface $\mathcal Z$ and of rate of crossing $\mathcal Z$
in direction from initial to final state \cite{ref6,ref10}. 
It can be written as

\begin{equation}\label{eq15}
\kappa=\frac 1{\tau }=Z^{-1}\int_{v^\perp>0}v^\perp(\mathbf \Delta)e^{-\frac{E[\mathbf S]}{k_BT}}d\mathbf \Delta,
\end{equation}
where  $v^\perp$ is the component of velocity 
${d\mathbf S}/{dt}$ orthogonal to the dividing surface $\mathcal Z$
and integral is taken over a part of $\mathcal Z$ where $v^\perp>0$. 
For simplicity we suppose that $\mathcal Z$ is plane and eigenvector $\mathbf e_{TS}^{(1)}$ 
is orthogonal to $\mathcal Z$ at the transition state. 
The value $v^\perp$ can be calculated using the Landau-Lifshitz equation in the harmonic approximation. 
In a neighborhood of saddle point the equation can be linearized
resulting in equation of motion in the tangent space:

\begin{equation}\label{eq:LLL}
\frac d{dt}\mathbf\Delta_n=\frac{\gamma }{\mu_n} \left[ \mathbf S_n^{TS} \times \sum _{\langle n,m\rangle} 
\mathcal H_{nm}[\mathbf S^{TS}] \mathbf\Delta _m \right].
\end{equation}
The component of the velocity orthogonal to the dividing surface is given by

\begin{equation*}
v^\perp=\sum_n \mathbf e_{TS}^{(1)}\cdot\frac{d\mathbf\Delta _n}{dt}
=\mathbf e_{TS}^{(1)} \cdot \sum_n \frac{\gamma }{\mu_n} \left[ \mathbf S_n^{TS} \times 
\sum _{<n,m>} \mathcal H_{nm}[\mathbf S^{TS}] \mathbf\Delta_m \right].
\end{equation*}
Decomposing the increments $\mathbf \Delta$ over the eigenvectors of Hessian $\mathcal H$ 
(that is $\mathbf \Delta=\sum_i\Delta_{TS}^{(i)}\mathbf e_{TS}^{(i)}$),
we have:

\begin{equation}\label{eq18}
v^\perp=\sum_i \mathbf e_{TS}^{(1)}\cdot\zeta _{TS}^{(i)}\Delta_{TS}^{(i)}
\sum_m\frac{\gamma }{\mu _m}[\mathbf S_m^{TS}\times \mathbf e_{TS}^{(n)}]
\equiv
\sum_i a_{TS}^{(i)}\Delta_{TS}^{(i)},
\end{equation}
where  
$a_{TS}^{(i)}=\mathbf e_{TS}^{(1)}\cdot\sum_n\frac{\gamma \zeta_{TS}^{(i)}}{\mu_n}
[\mathbf S_n^{TS}\times \mathbf e_{TS}^{(i)}]$
shows how fast the perpendicular component of velocity  $v^\perp$ increases when moving from a stationary point 
$\mathbf S^{TS}$ in the direction of the eigenvector of the Hessian $\mathbf e_{TS}^{(i)}$. 
Clearly $a_{TS}^{(1)}=0$. 
Now transition rate can be written as

\begin{equation}\label{eq19}
\kappa=Z^{-1}e^{-\frac{E[\mathbf S^{TS}]}{k_BT}}\int_{\mathcal Z'}
\sum_i a_{TS}^{(i)}\mathbf \Delta_{TS}^{(i)}\exp \left(-\frac{\sum_j\zeta_{TS}^{(j)}
(\mathbf \Delta_{TS}^{(j)})^2}{k_BT}\right)d\mathbf \Delta,
\end{equation}
here the integral is taken over part $\mathcal Z'$ of the dividing surface 
$\mathcal Z$ (i.e. $a_{TS}^{(1)}=0$) 
where $\sum_na_{TS}^{(i)}\mathbf \Delta_{TS}^{(i)}>0$.

The expression in the brackets is the square of the length of a vector in 
the $2N$-dimensional configuration space. 
This length is expressed in terms of the coordinates 
$\mathbf \Delta_{TS} ^ {(i)} \sqrt {\frac {\zeta_{TS}^{(i)}}{k_BT}} $ 
in the basis $ \{\mathbf e_ {TS}^{(i)}\}$, but can be written in any orthonormal basis. 
If we choose one of the vectors of this basis along $\mathbf v^\perp$, 
then the boundaries of integration area will have very simple form, 
and all integrals can be calculated explicitly. 
As result we will have for rate:

\begin{equation}\label{eq23}
\kappa=Ce^{-\frac{E[\mathbf S^{TS}]}{k_BT}}\frac{({\pi k}_BT)^N}{2\pi } \left(\prod _{l=2}^{2N}
\zeta_{TS}^{(l)}\right)^{-1/2}\sqrt{\sum_j\frac{(a_{TS}^{(j)})^2}{\zeta_{TS}^{(j)}}}
\end{equation}
Substituting expression (\ref{eq14}) gives: 

\begin{equation}\label{eq24}
\kappa=\frac 1{2\pi }\sqrt{\frac{\det\mathcal H^m}{\prod _{l=2}^{2N}\zeta_{TS}^{(l)}}}
\sqrt{\sum_j\frac{(a_{TS}^{(j)})^2}{\zeta_{TS}^{(j)}}}e^{\frac{E[\mathbf S^m]-E[\mathbf S^{TS}]}{k_BT}}
\end{equation}

This expression exactly coincides with one reported in the ref. \cite{ref6}
if $\det \mathcal H^m $ is written as product of the eigenvalues $\zeta_m^{(i)}$.

The states on the MEP near TS can be expressed as
$\mathbf S=\mathbf S^{TS}+\varepsilon \mathbf e_{TS}^{(1)}+o(\varepsilon)$.
These states evolve according to linearized 
Landau-Lifschitz equation (\ref{eq:LLL}) as follows:
\begin{equation}\label{eq25}
\frac{1}{\varepsilon}\frac{d \mathbf S_n}{dt}+o(1)
=\zeta^{(1)}_{TS}
\sum_n\frac{\gamma }{\mu_n}[\mathbf S_n^{TS}\times \mathbf e_{TS,n}^{(1)}]
\equiv
\frac{d \mathbf e_{TS,n}^{(1)}}{dt}.
\end{equation}
The vector $d\mathbf e_{TS}^{(1)}/dt$ defined by the identity (\ref{eq25})
is the velocity of the state $\mathbf S$ shifted from TS by an infinitesimal distance $\varepsilon$
in the direction $\mathbf e_{TS}^{(1)}$
divided by $\varepsilon$.
Denote by $b^{(i)}$ the projection of velocity 
$d \mathbf e_{TS}^{(1)}/dt$ along the MEP into eigensubspace 
$\mathbf e_{TS}^{(i)}$ of $\widehat  H[\mathbf S^{TS}]$.

\begin{equation}\label{eq26}
b^{(i)}=\mathbf e_{TS}^{(i)}\cdot\frac{d \mathbf e_{TS}^{(1)}}{dt}
=\frac{\gamma }{\mu}\zeta_{TS}^{(1)}\mathbf e_{TS}^{(i)}\cdot[ 
  \mathbf S^{TS}\times \mathbf e_{TS}^{(1)}]
=\frac{\zeta_{TS}^{(1)}}{\zeta _{TS}^{(i)}}a_{TS}^{(i)}
\end{equation}

It is clear that the vectors $\mathbf b$ and $ \mathbf a$ are in one-to-one correspondence, but the vector $\mathbf a$ is introduced as a characteristic of the dividing surface, whereas the vector $\mathbf b$ depends only on the motion
of the magnetic state along the MEP.

Now the expression for rate can be written in an alternative form:

\begin{equation}\label{eq27}
\kappa=\frac 1{2\pi }\sqrt{\frac{\det\mathcal H^m}{|\det\mathcal H^{TS}|}}
\sqrt{\sum_{i=2}^{2N}(b^i)^2\zeta_{TS}^{(i)}}
\exp\frac{E[\mathbf S^m]-E[\mathbf S^{TS}]}{k_BT}
\end{equation}
The sum over eigenvalues can be written as quadratic form with matrix  $\mathcal H^{TS}$ on vector $\mathbf b$:

\begin{equation}\label{eq:dyn}
\sum_{i=2}^{2N}(b^i)^2\zeta _{TS}^{(i)}=\mathbf b\cdot\mathcal H^{TS} \mathbf b
\end{equation}
It can be computed in arbitrary basis, including initial basis of individual spins. 
Therefore eigenvalue decomposition is no longer necessary, 
that give opportunity to speed up numerical computation of transition rate. 
Such method was used in ref. \cite{ref28} for calculation of Sk lifetime in antiferromagnet. 

Finally, the transition rate in HTST is computed as follows:

\begin{equation}\label{eq28}
\kappa=\frac{\kappa^{ent}\kappa^{dyn}}{2\pi}
\exp{\frac{E[\mathbf S^m]-E[\mathbf S^{TS}]}{k_BT}},
\end{equation}
where the entropy factor $\kappa^{ent}$ is determined by free energies at state in consideration,
and dynamical factor $\kappa^{dyn}$ characterizes motion about TS: 
\begin{equation*}
\kappa^{ent}=\sqrt{\frac{\det\mathcal H^m}{|\det\mathcal H^{ts}|}},\quad
\kappa^{dyn}=\sqrt{\mathbf b\cdot\mathcal H^{ts} \mathbf b},
\end{equation*}
here $\mathbf b$ is unit tangent vector to the MEP at the saddle point.

The expression for transition rate within HTST presented above takes into account all
vibrational modes and count all trajectories in the neighborhood of transition state for which the velocity is directed
toward the final state. 
This approach can work for relatively large damping and has to be corrected via coefficient
which take into account recrossing of dividing surface. 

Similar expression for the rate of magnetic transitions can be obtained within the
Kramers-Langer theory \cite{ref8}

\begin{equation}\label{eq29}
\kappa=\frac{\lambda _{+}\kappa^{ent}}{2\pi }
\exp{\frac{E[\mathbf S^m]-E[\mathbf S^{TS}]}{k_BT}}
\end{equation}
where $\lambda_+$ is the positive eigenvalue of the operator corresponding 
the linearized Landau-Lifshitz-Gilbert equation (\ref{eq:LLL}). 
Only dynamical parts in eqs. (\ref{eq28}) and (\ref{eq29}) are different. 
In both HTST and Kramers-Langer's theory the most
computationally expensive part is calculation  determinants of Hessian of energy. 
Below we develop methods to speed up computations of determinants, 
which allow analysis of structures with millions of
magnetic moments.
 
\section{Block LU decomposition and determinant.}
\label{sec:detlu}

When calculating the rate of magnetic transitions (\ref{eq28}), the most difficult task from a computational point of view is to calculate the determinants of the Hessian of 
energy for systems with huge number of degrees of freedom, which is usually the case for topologically protected structures. 
If the initial configurations $\mathbf S^m$ is given and its energy $E[\mathbf S^m]$ is known, the first step is to find the saddle points on the energy surface corresponding to the 
transition state $\mathbf S^{TS}$. A number of methods have been developed for determining saddle points and constructing an MEP between the initial and final magnetic configurations. 
For chiral magnetic structures, such as magnetic Sks, one can use the Geodesic Nudged Elastic Band (GNEB) method \cite{ref9}, the string method \cite{Heistracher18}, the truncated MEP method \cite{ref25}, which only find a part of the MEP in the vicinity of TS for large systems with millions moments and a Minimum Mode Following (MMF) method \cite{Muller18} that can be used without knowing the final state and predict unknown transition scenarios.

When TS is found, the computation of dynamic prefactor (\ref{eq:dyn}) 
is as expensive as single computation of energy.
The entropy prefactor contains the determinants of Hessian of energy. It is  the most challenging value to compute due to the large dimensionality  of the problem. Without exploiting peculiarities of the Hessian matrices these calculations are  possible for system with the number of magnetic moments up to $10^5$.

One way to overcome the difficulty is statistical evaluation,
see for example \cite {PW18, MG07, BDKKZ17},
however, we are not aware of the successful implementation
method to magnetic systems.

Another possibility is to exploit the fact that the Hessian matrices in the initial and transition states have very similar eigenvalues, and only part of the ratio of the eigenvalues differs significantly from unity, see, for example, \cite {ref16}.
However, in the benchmark section below, we perform a computational experiment showing that almost all eigenvalues must be computed in order to obtain sufficient prefactor accuracy.

Below we propose a method for calculating determinants using the sparsity of the Hessian matrix for a system without long-range interactions, such as dipole interaction.
Although the dipole interaction plays an important role in the stabilization 
of large-scale Sks and other topological structures \cite{ref5}, as a first 
approximation this interaction can be taken into account using the effective anisotropy \cite{ref14}.  Such approach fits perfectly into the proposed scheme.

Hessian of energy in Heisenberg-like model is given by the equation (\ref{eq:projhess}).
The Hessian for a state $S^0$ is $2N\times 2N$ matrix written in the form 
$\mathcal H=\Pi^T\widehat H\Pi$,
where $\widehat H$ is $3N\times 3N$ matrix of the quadratic form of energy minus 
a diagonal matrix providing corrections on curvature of the constraints manifold,
$\Pi$ is an embedding of $\mathbb R^{2N}$ to the tangent space to the manifold at $\mathbf S^0$.
It is convenient to consider the matrix $\widehat H$ as a block matrix with $3\times3$ blocks 
$\widehat H_{nm}$ describing interaction of spins $n$ and $m$.
The $ \widehat H$ matrix is sparse if the interaction is short-range and 
only a few neighbors need to be taken into account for each magnetic moment.

Since the embedding $\Pi$ can be naturally defined as the  direct sum of embedding on each sphere for each spin as in (\ref{eq:tbasis}),
the matrix $\mathcal H$ has the same sparse structure as $\widehat H$.
The choice $\Pi$ is not unique, but determinant of $H$ does not depend on $\Pi$.
It is possible to eliminate $\Pi$ from computations completely, 
if we notice that $\mathcal H$ has the same determinant as the matrix $P\widehat H P+(1-P)$
having again the same sparse structure as $\widehat H$,
where the matrix $P$ is a projector to the tangent space $P=\Pi\Pi^T$, 
which can be written in terms of $\mathbf S^0$:

\[
  P=\diag_n P_n,\quad P_n=1-|\mathbf S^0_n\rangle \langle \mathbf S^0_n|.
\]  
The simplification is however comes at the cost of computing determinant of $3N\times 3N$ matrix
instead of $2N\times 2N$.

The indices $n$, $m$ for matrix elements $H_{nm}$ are in fact a tuple 
of coordinates of spins, e.g. $n=(n_1, n_2, n_3)$,
where $H_{nm}$ acts on spin $\mathbf S_m$.
It is convenient to group spins into layers $\tilde S_{n_1}=(\mathbf S_{n_1,n_2,n_3})_{n_2,n_3}$
containing spins having the same first index.
The elements of $H$ can also be regrouped into blocks $\tilde H_{n_1,m_1}$ 
acting on the layers $\tilde S_{m_1}$. 
Having system of size $N=N_1N_2N_3$, $n_k=1\ldots N_k$,
the blocks $\tilde H_{n_1, m_1}$ are of size $M\times M$, where $M=2N_2N_3$.
Sufficiently separated layers $n_1$ and $m_1$ do not interact, 
therefore, the block matrix $\tilde H$ has few non zero diagonals (banded matrix). 
The matrix can be transformed to block triangular form doing LU or QR decomposition.
Then calculation of the determinant of $H$ is simplified to $N_1$ computations of the determinants of the diagonal blocks of the triangular matrix.
Below we demonstrate how this can be done for nearest neighbors model, where the Hessian matrix is block tridiagonal
with possible nonzero corners $\tilde B_{N_1}$ for periodic boundary conditions along the first axis:
\begin{equation}\label{eq30}
\tilde H=\begin{pmatrix}\tilde A_{1}&\tilde B_{1}^T&0&\cdots&\tilde B_{N_1}\\
\tilde B_{1}&\tilde A_{2}&\tilde B_{2}^T&\cdots&0\\
0&\tilde B_{2}&\tilde A_{3}&\cdots&0\\
\cdots & \cdots & \cdots & \ddots & \vdots\\
\tilde B_{N_1}^T & 0 & 0 & \cdots & \tilde A_{N_1}\\
\end{pmatrix}
\end{equation}
where $\tilde A_n=\tilde H_{nn}$, $\tilde B_n=\tilde H_{n+1,n}$.
Since the matrices $\tilde A_n$ and $\tilde B_m$ do not commute in general case,
the Leibniz formula for determinants is inapplicable here.
However block LU decomposition can be used to transform the matrix into 
an upper triangular form.
For a block matrix $ 2 $ by $ 2 $, the transformation leads to  Schur's determinant identity:

\[
\begin{vmatrix}
  \widehat  A&\widehat  B\\
  \widehat  C&\widehat D
\end{vmatrix}
=\begin{vmatrix}
  \widehat  A&\widehat  B\\
  0&\widehat  D-\widehat  C\widehat A^{-1}\widehat  B\,
\end{vmatrix}
=|\widehat  A|\cdot|\,\widehat  D-\widehat  C\widehat A^{-1}\widehat  B\,|,
\]
where according to the Gaussian decomposition procedure
the first row is multiplied by $-\widehat C\widehat  A^{-1}$ from the left and added to the second row,
and we used the fact that the determinant of the triangular block matrix is equal to the product of the determinants of the diagonal blocks.
In the same way we can transform the matrix $\tilde H$ into a similar block triangular matrix.

First we consider the free boundary conditions along the first axis,
that is, the matrix is block tridiagonal and corners vanish $\tilde B_{N_1}=0$.
After Gaussian decomposition we obtain an upper triangular matrix 
with the main diagonal consisting of elements $A_n'$,
where
\begin{equation}\label{eq:Arec}
  A_1'=\tilde A_1,\quad A_{n+1}'=\tilde A_{n+1}-\tilde B_n(A_{n}')^{-1}\tilde B_n^T.
\end{equation}
Since the determinant of the matrix itself is too large to be stored as a floating point number, we will use its logarithm.
The determinant in the transition state is negative, but only the absolute value of the determinant is required for TST, so we calculate the modulus of the determinant:

\begin{equation}\label{eq:detfree}
  \ln\det |\mathcal H|=\sum_{n=1}^{N_1} \ln\det |A_n'|.
\end{equation}

Here the determinant can be calculated using the recursive formula (\ref{eq:Arec}) in $O(N_1)$ time, which gives a significant improvement over the $O(N_1^3)$ complexity for a dense matrix.
For the Hessian matrix containing $ K $ off-diagonal blocks, the complexity of the block LU  decomposition is $O(N_1K^2)$, which remains linear with respect to $N_1$.
However, regarding the block size $M$, the algorithm has a complexity of $O(M^3)$ (or less for fast matrix multiplication), since we still need to invert the diagonal elements of $A_n'$, which are dense matrices for large $N$. This means that for maximum performance, one should use block LU decomposition with respect to the largest axis $N_k$.

The $H$ matrix discussed in this section depends on the  matrix $\widehat H$ of quadratic form of energy as well as the embeddings $\Pi_n$. For uniform media, all matrix elements of $\widehat H$ on one diagonal are the same, but for non-uniform states the embeddings $\Pi_n$ are, in general case, different. Therefore the matrix elements $\tilde A_n$ and $\tilde B_n$ are of a general form.
However, for the uniformly magnetized state (UMS) $\mathbf S^0$ (and other states where $\mathbf S^0_n$ does not depend on $n_1$, for example, the domain wall along the $\mathbf x$ axis) all diagonals $H$ are constant $\tilde A_n=\tilde A_1$, $\tilde B_n=\tilde B_1$. In this case, the evaluation of $A_n'$ is a simple iteration of the function $f(x) = \tilde A_1 - \tilde B_1 x^{-1}\tilde B_1^T$, acting on symmetric matrices $x$. Assuming $f$ is a contraction map, we conclude that $\tilde A_*'$ converges fast to the fixed point $\tilde A_*'$. Then 
\[
  \ln \det H=N_1\ln\det A_*'[S^0] + O(1),\text{ as }N_1\to\infty.
\]
For the metastable state $\mathbf S^m$ and the transition state $\mathbf S^{TS}$, the entropy prefactor for large $N_1$ can be approximated as follows:
\[
  \ln k^{ent} 
  = \frac12 N_1(\ln\det A_*'[\mathbf S^m]-\ln\det A_*'[\mathbf S^{TS}])+O(1),
\]
as $N_1\to\infty$,
that is, one dimension can be removed from the analysis.
This approach makes it easy to obtain the lifetimes of states obtained by repeating $N_1$ times of single layer of the system. 
It is worth noting that interlayer interaction is still present in the equation rate, and $A_*'$ does not coincide with $A_1$, and the pre-exponential factor exponentially depends on the number of layers $N_1$.

For periodic boundary conditions and short-range interaction, the determinant can also be calculated using the recursive formula, where $A'_n$ is the same as above, and
\[
W_1 = \tilde B_{N_1},\quad W_{n+1} = \tilde B_{n+1}^T\delta_{n+1,N_1-1} - \tilde B_n (A_{n}')^{-1} W_{n},
\]
\[
Q_1 = A_{N_1},\quad Q_{n+1} = Q_{n} - W_{n}^T(A_{n}')^{-1}W_{n}.
\]
The log-det in this case is computed as
\begin{equation}\label{eq:detper}
  \ln\det |\mathcal H|=\ln\det|Q_{N_1}|+\sum_{i=1}^{N_1-1} \ln\det |A_i'|.  
\end{equation}

Alternatively, the determinant of a matrix with corners or any other perturbation can be calculated using the matrix determinant lemma

\begin{equation}\label{eq:detlem}
\det(F + U V^T) = \det(1 + V^T F^{-1}U)\cdot\det F,
\end{equation}
and Woodbury matrix identity:
\begin{equation}\label{eq:woodbury}
  (F + U V^T)^{-1} = F^{-1} - F^{-1}U(1 + V^T F^{-1} U)^{-1}V^TF^{-1}.
\end{equation}
For examples, denoting by $F$ Hessian for free boundary conditions, 
the Hessian matrix $H$ for periodic boundary conditions can be written 
as $H=F+UV^T+VU^T$ in terms of two block vectors:
\[
  U=\begin{pmatrix}1&0&\ldots&0\end{pmatrix}^T,\quad
  V=\begin{pmatrix}0&\ldots&0&B_{N_1}\end{pmatrix}^T.
\]
Using (\ref{eq:detlem}) and (\ref{eq:woodbury}) we obtain:  
\[
\det \mathcal H
= \det(1 + a^T - b(1+a)^{-1}c)\det(1+a)\det F,
\]
where $a=V^TF^{-1}U$, $b=U^TF^{-1}U$, $c=V^TF^{-1}V$. The computation of the first two determinants is relatively fast, since they are determinants of small matrices $M\times M$. The matrix $F$ is tridiagonal, therefore, one can find the value  $x=F^{-1}U$ as a solution of the equation $Fx=U$ using Gaussian elimination as above. The same is true for $F^{-1}V$. 

The matrices $a$, $b$ and $c$ can be found explicitly by the following identities:
\[
  a = B_{N_1}(A_{N_1}')^{-1}D_{N_1},\quad
  b = (A_1')^{-1}T_1,\quad
  c=B_{N_1}^T(A_{N_1}')^{-1}B_{N_1},
\]
\[
  D_1=1,\quad D_{n+1}=-B_n(A_n')^{-1}D_n,
\]
\[
  T_{N_1}=D_{N_1},\quad T_{n-1}=D_{n-1}-B_{n-1}^T(A_n')^{-1}T_n
\]
where the recursion is first applied upward for $D$, 
and after that downward for $T$.


\section{Benchmark}
 
\begin{figure*}[t]
  \centering
    \includegraphics[width=0.49\textwidth]{"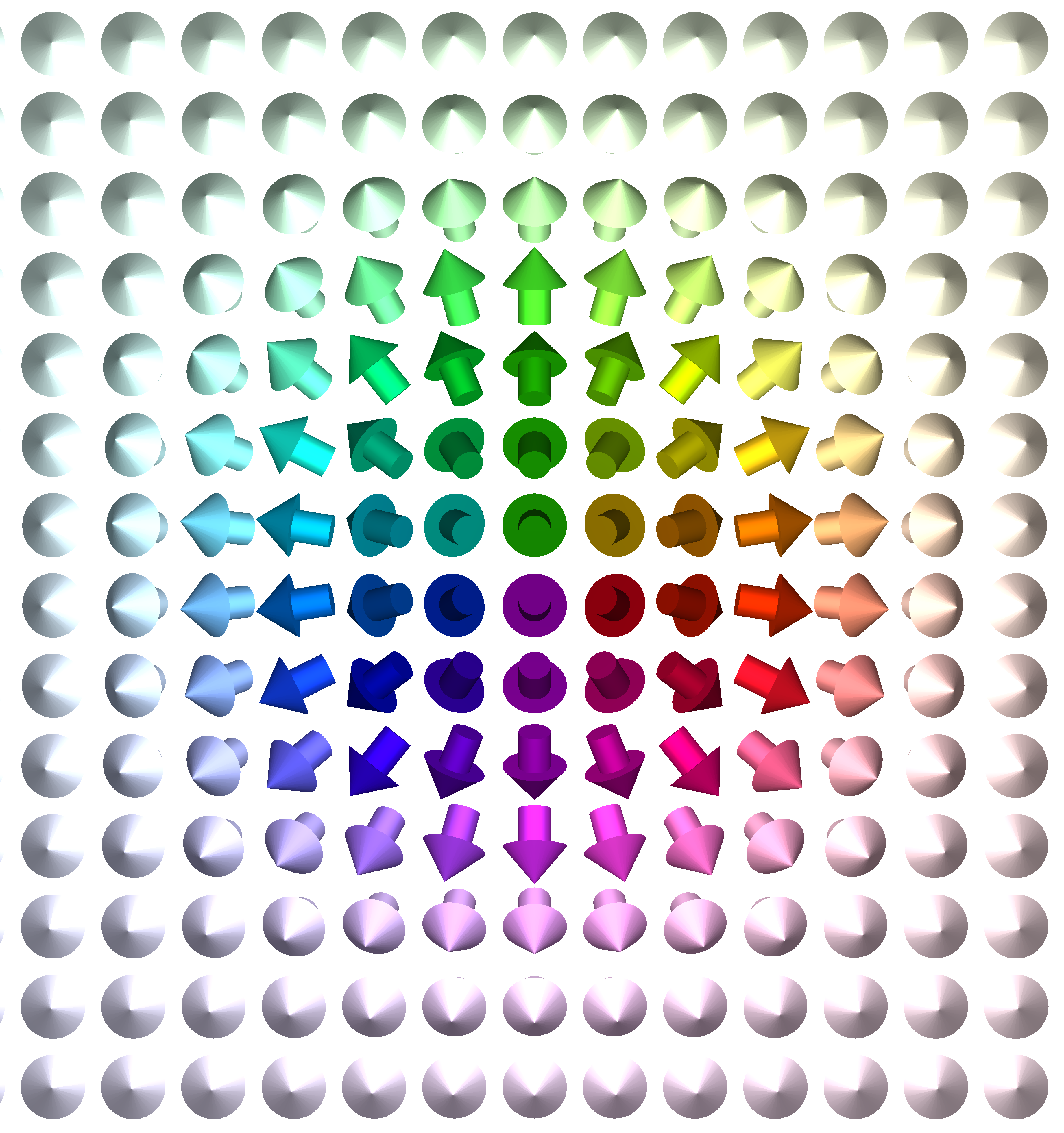"}
    \hfill
    \includegraphics[width=0.49\textwidth]{"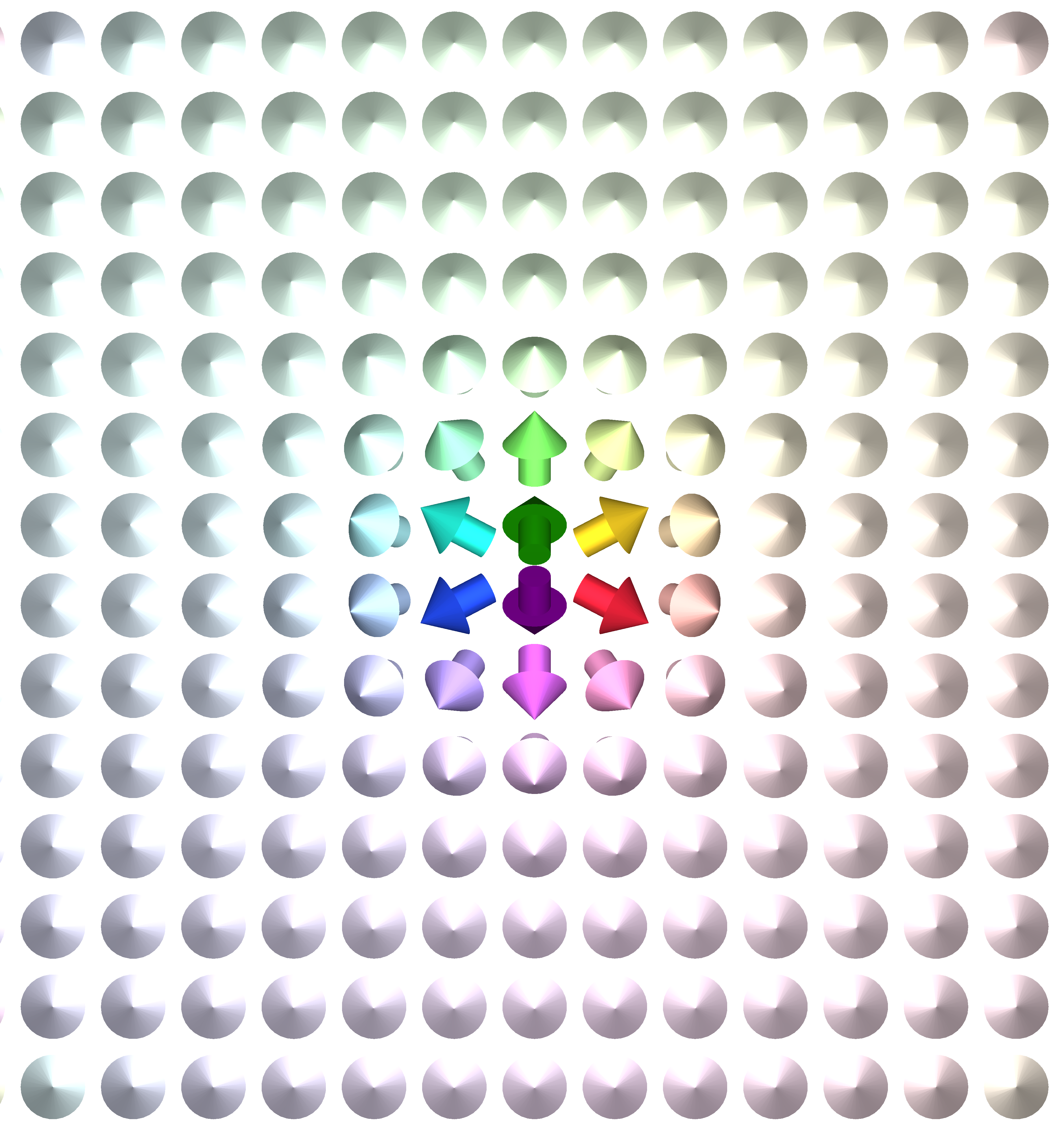"}
    \\[5mm]
    \includegraphics[width=0.49\textwidth]{"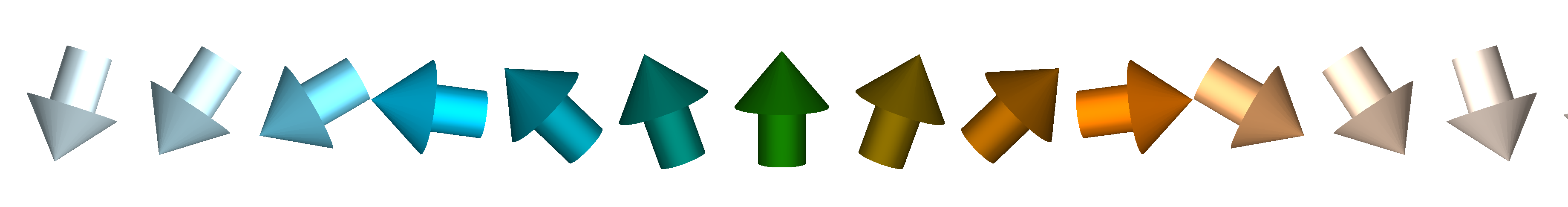"}
    \hfill
    \includegraphics[width=0.49\textwidth]{"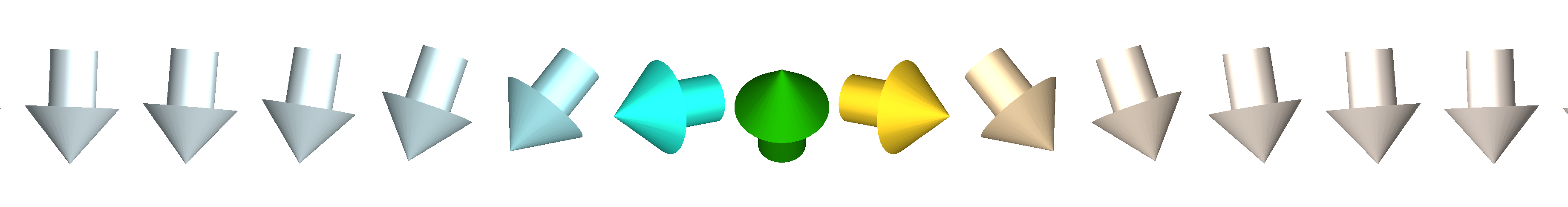"}

  \caption{
  Central part of the lattice containing isolated equilibrium Sk (left) and TS for decay in the benchmark [2P] (right)
  for the smallest lattice $30\times 31$:
  (upper panel) top view, (bottom panel) side view of 1D slice containing the center of the Sk.
  Colors encode spin orientation.
  }
  \label{fig:bench2}
\end{figure*}

\begin{figure*}[t]
  \centering
    \includegraphics[width=0.49\textwidth]{"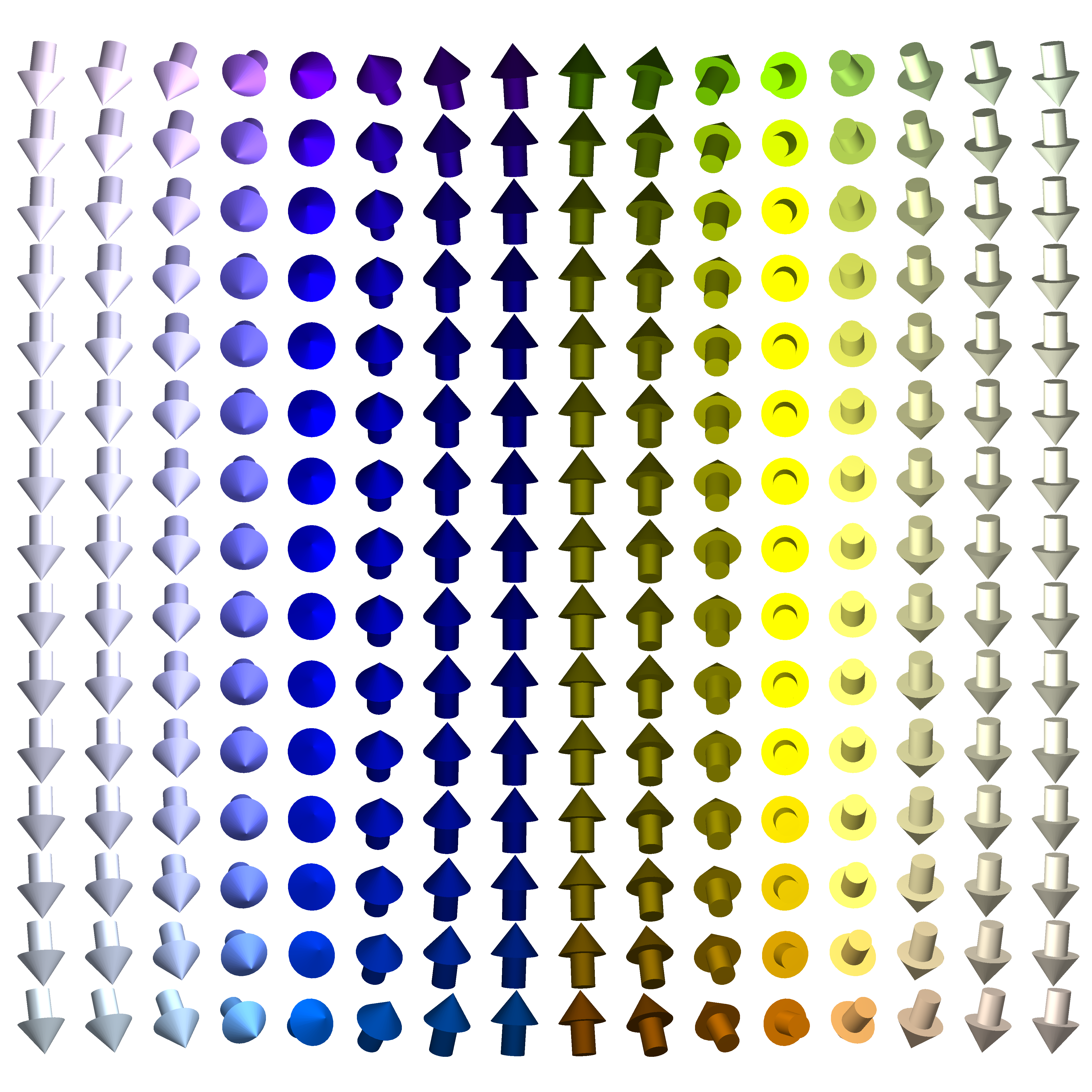"}
    \hfill
    \includegraphics[width=0.49\textwidth]{"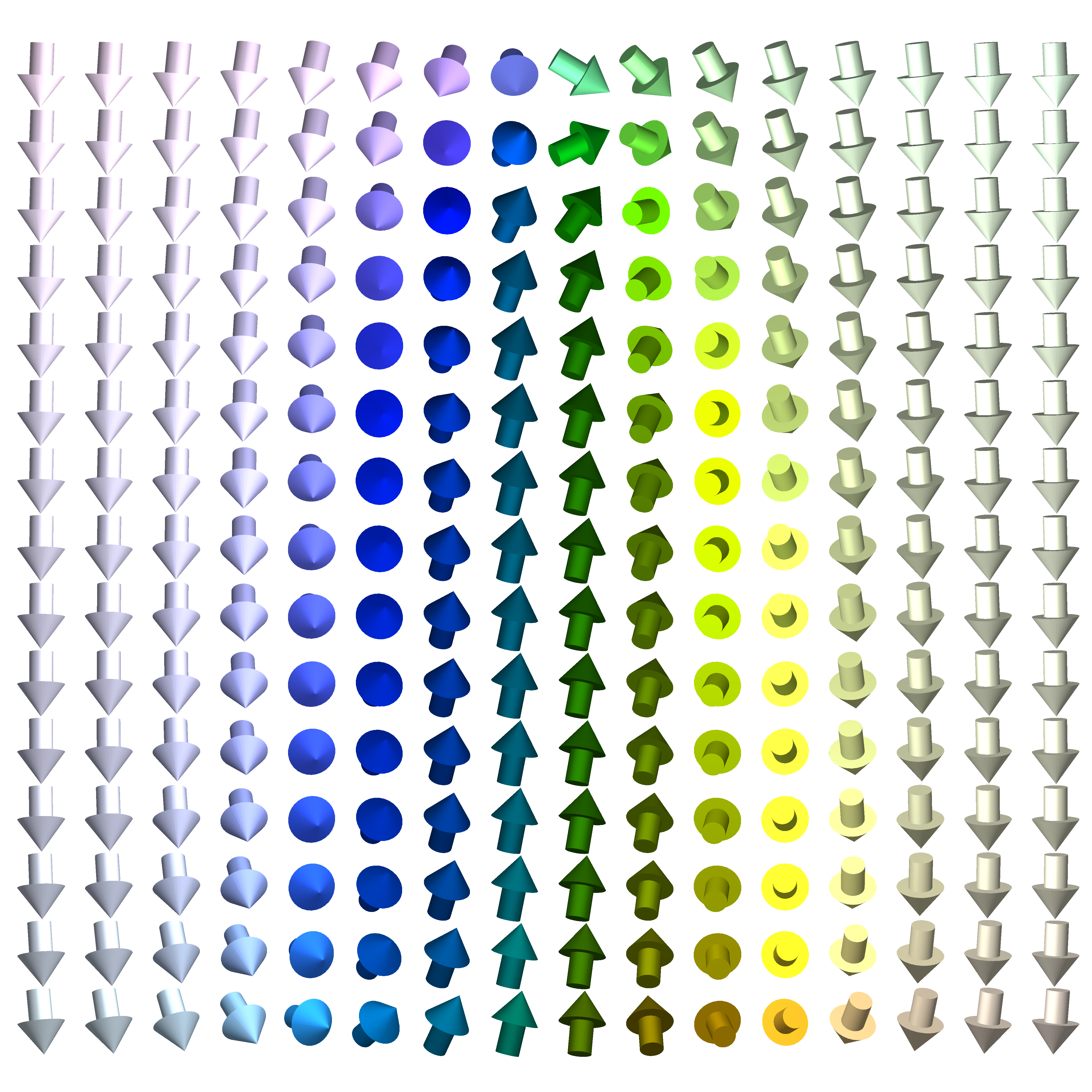"}
  \caption{
  A slice of three dimensional lattice through the axis of symmetry of Sk tube for the benchmark [3D]: (left) Sk tube state, (right)
  TS for annihilation process.
  The thickness $W$ of the sample is $15$ layers.
  Colors encode spin orientation.
  }
  \label{fig:bench3}
\end{figure*}

\begin{figure*}[t]
  \centering
  \begin{subfigure}{0.6\textwidth}
    \includegraphics[width=\textwidth]{"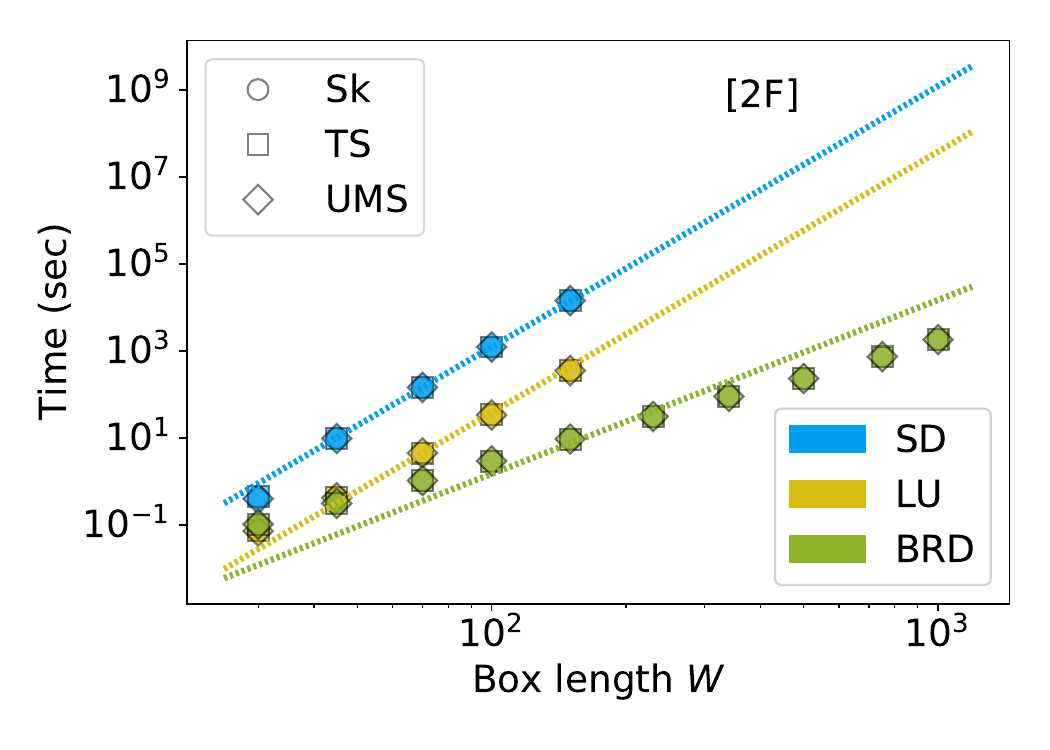"}
  \end{subfigure}
  \begin{subfigure}{0.6\textwidth}
    \includegraphics[width=\textwidth]{"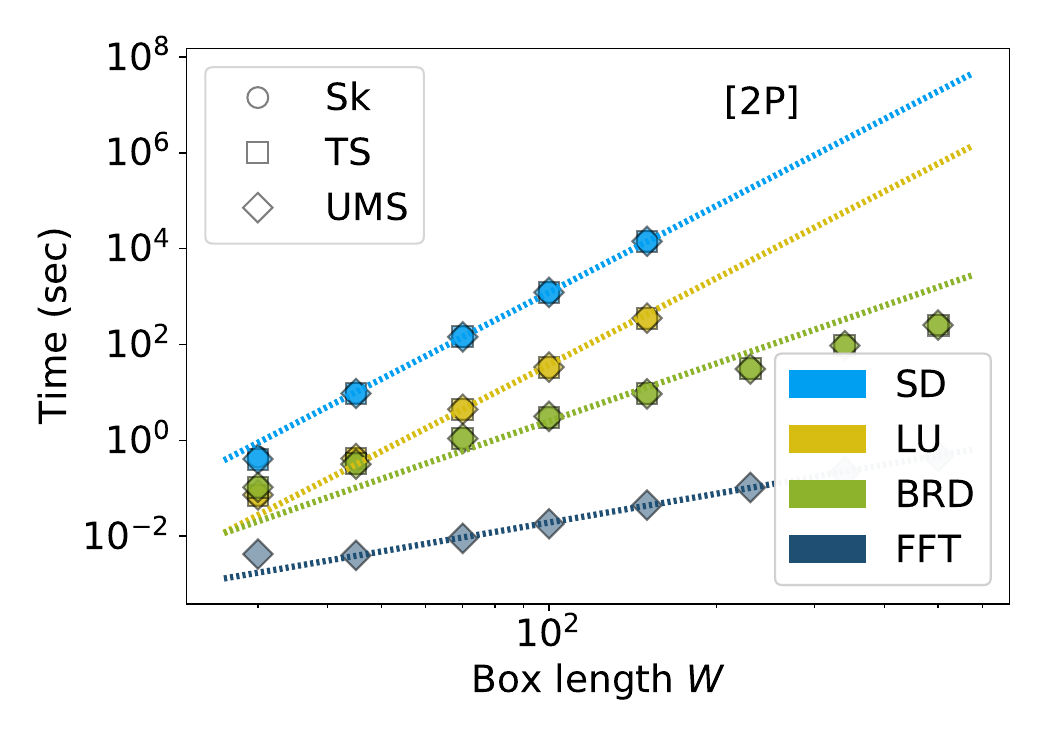"}
  \end{subfigure}
  \begin{subfigure}{0.6\textwidth}
    \includegraphics[width=\textwidth]{"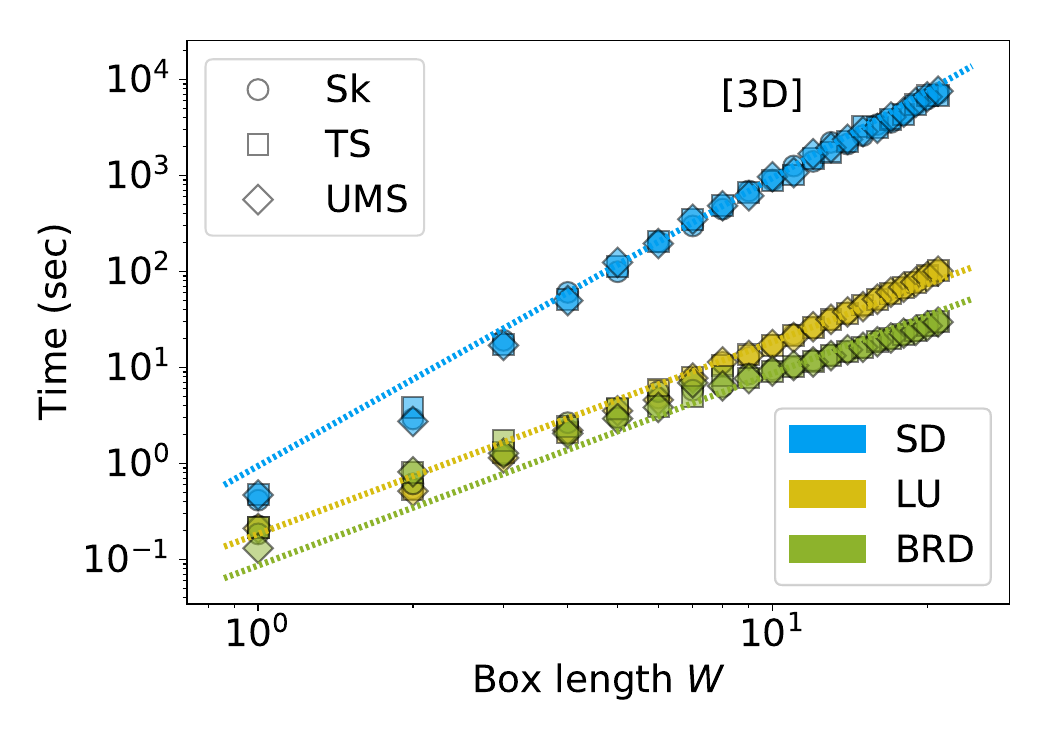"}
  \end{subfigure}
  \caption{
  Determinant computation time as function of box length $W$.
  Upper and middle panels: two dimensional lattice with free and periodic boundary conditions, respectively.
  Bottom panel: three dimensional system with fixed size in x-y plane and increasing thickness. 
  Test are performed for transition from Sk
  to UMS through TS.
  Points represent results of numerical experiments.
  Box sizes for [2F] and [2P] are given by Table 1, the box for [3D] case has size $30\times31\times W$.
  Theoretical predictions of computation time (\ref{eq:theortime}) are drawn as dotted lines.
  }
  \label{fig:time}
\end{figure*}

\begin{figure*}[t]
  \centering
  \begin{subfigure}{0.6\textwidth}
    \includegraphics[width=\textwidth]{"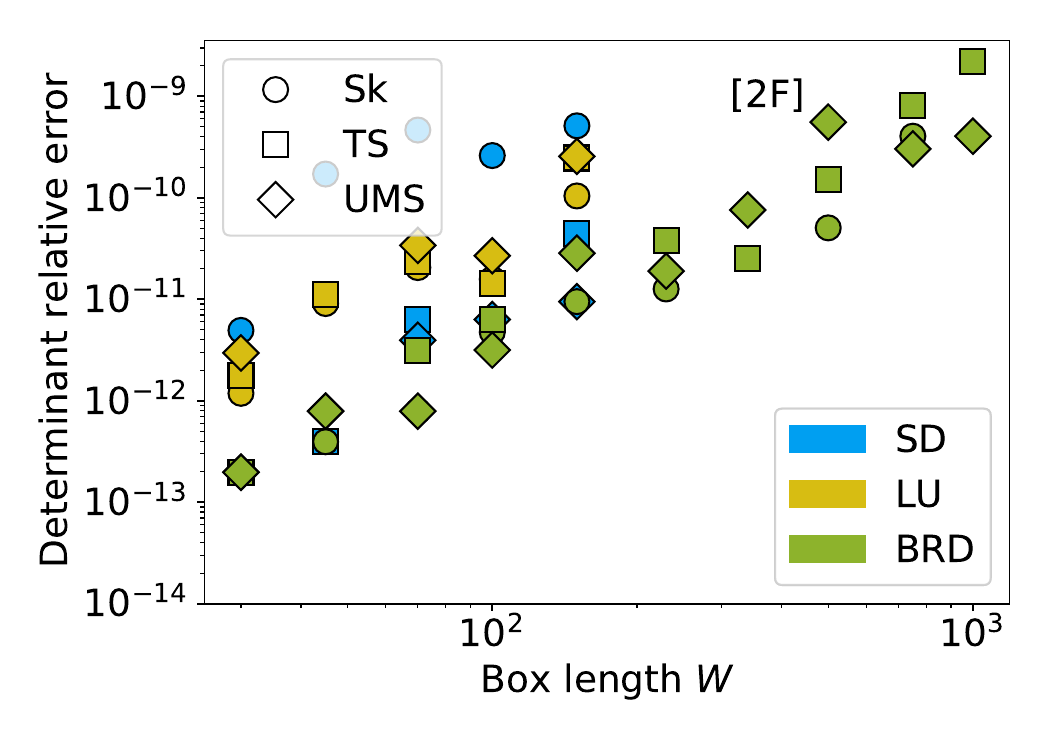"}
  \end{subfigure}
  \begin{subfigure}{0.6\textwidth}
    \includegraphics[width=\textwidth]{"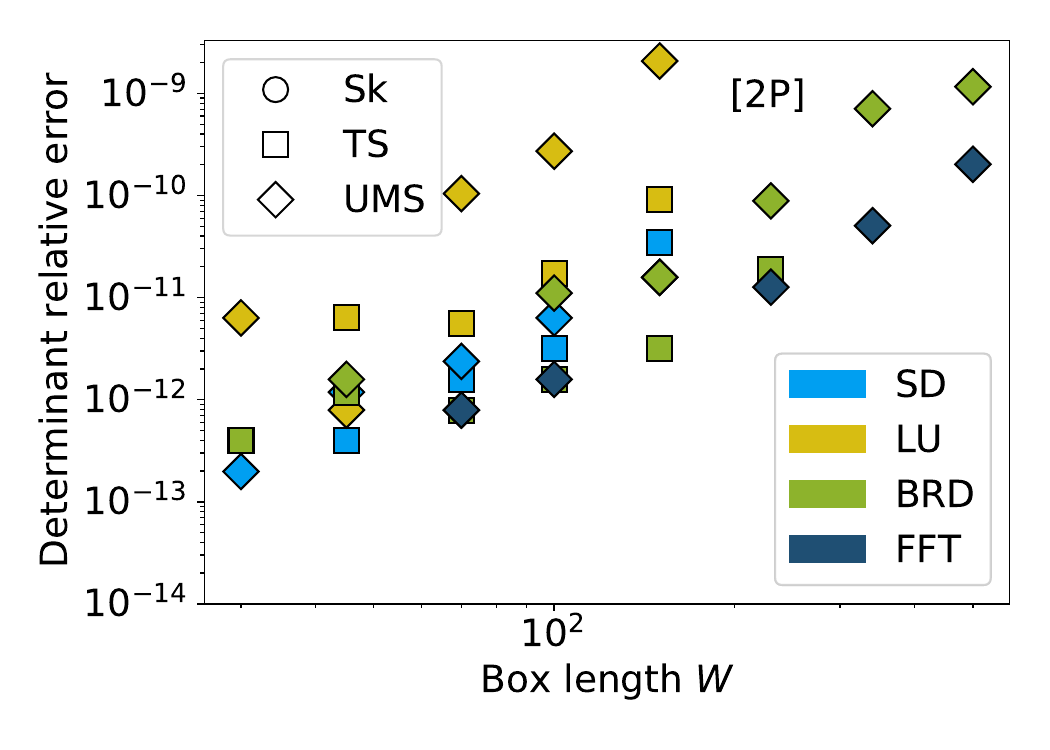"}
  \end{subfigure}
  \begin{subfigure}{0.6\textwidth}
    \includegraphics[width=\textwidth]{"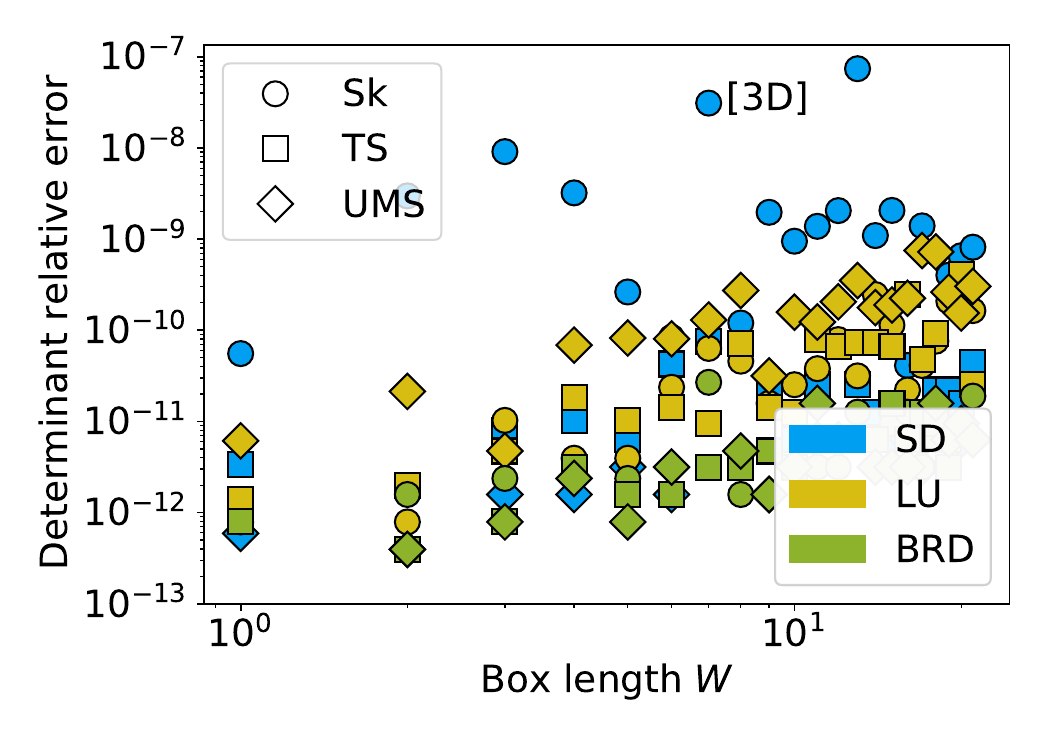"}
  \end{subfigure}
  \caption{Determinant computation precision as function of system size.
  Upper and middle panels: two dimensional lattice with free and periodic boundary conditions, respectively.
  Bottom panel: three dimensional system with fixed size in x-y plane and increasing thickness $W$. 
  Test are performed for transitions from Sk
  to UMS through TS.
  Box sizes are the same as in Fig. \ref{fig:time}.
  }
  \label{fig:det}
\end{figure*}

\begin{figure*}[t]
  \centering
  \begin{subfigure}{0.6\textwidth}
    \includegraphics[width=\textwidth]{"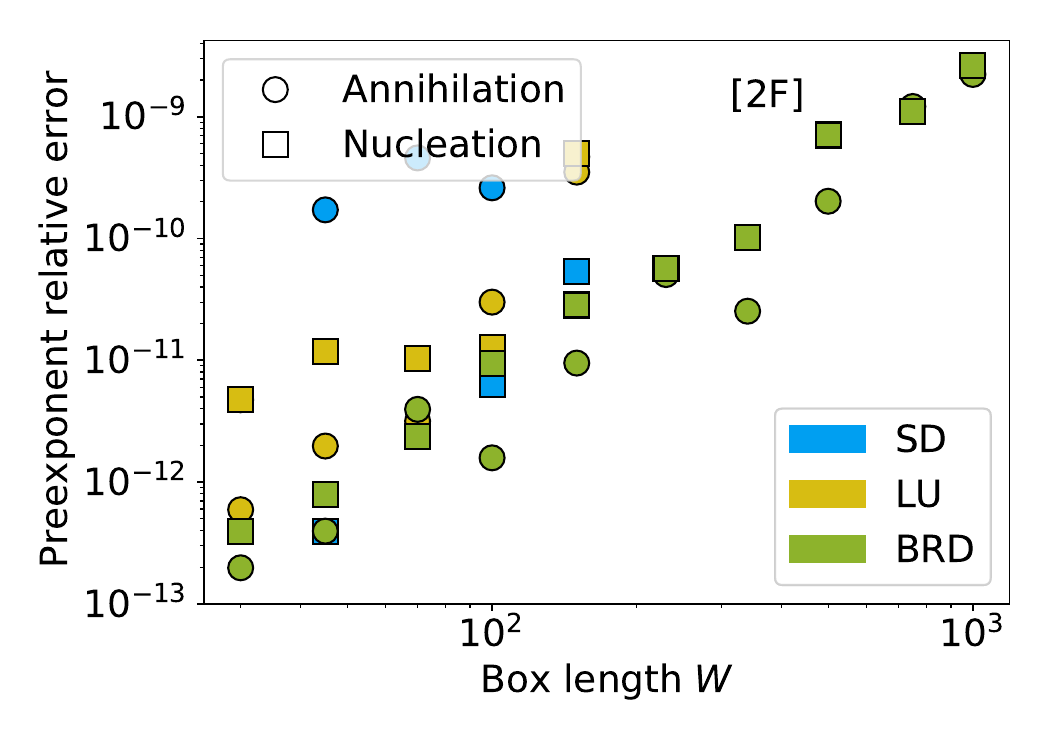"}
  \end{subfigure}
  \begin{subfigure}{0.6\textwidth}
    \includegraphics[width=\textwidth]{"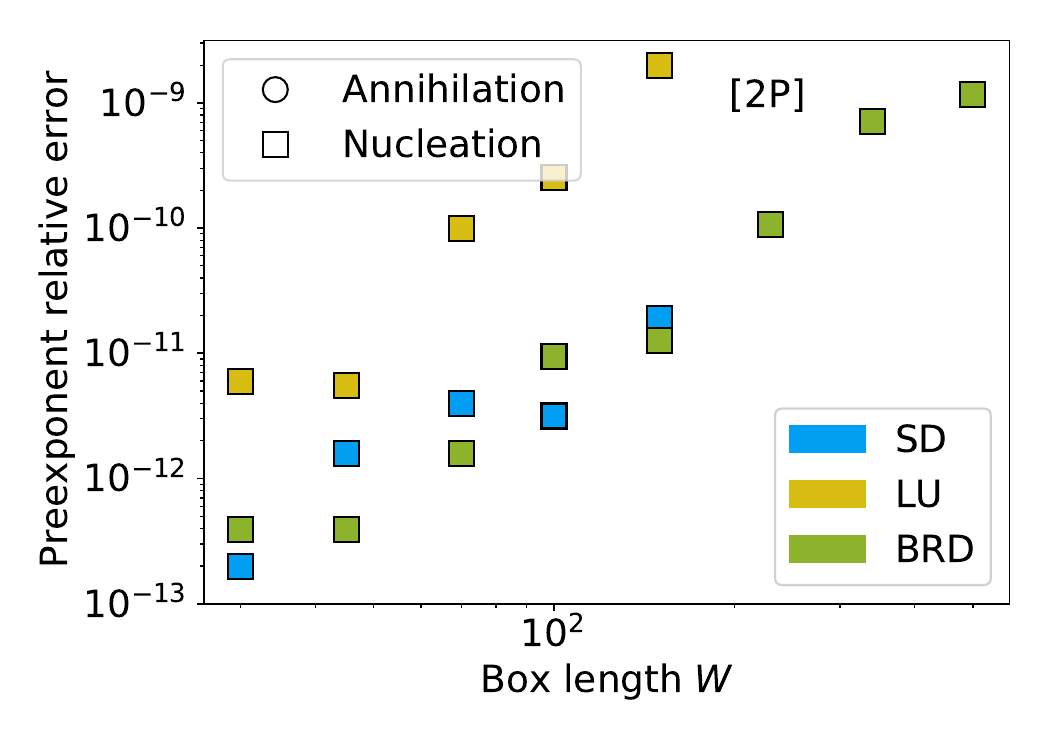"}
  \end{subfigure}
  \begin{subfigure}{0.6\textwidth}
    \includegraphics[width=\textwidth]{"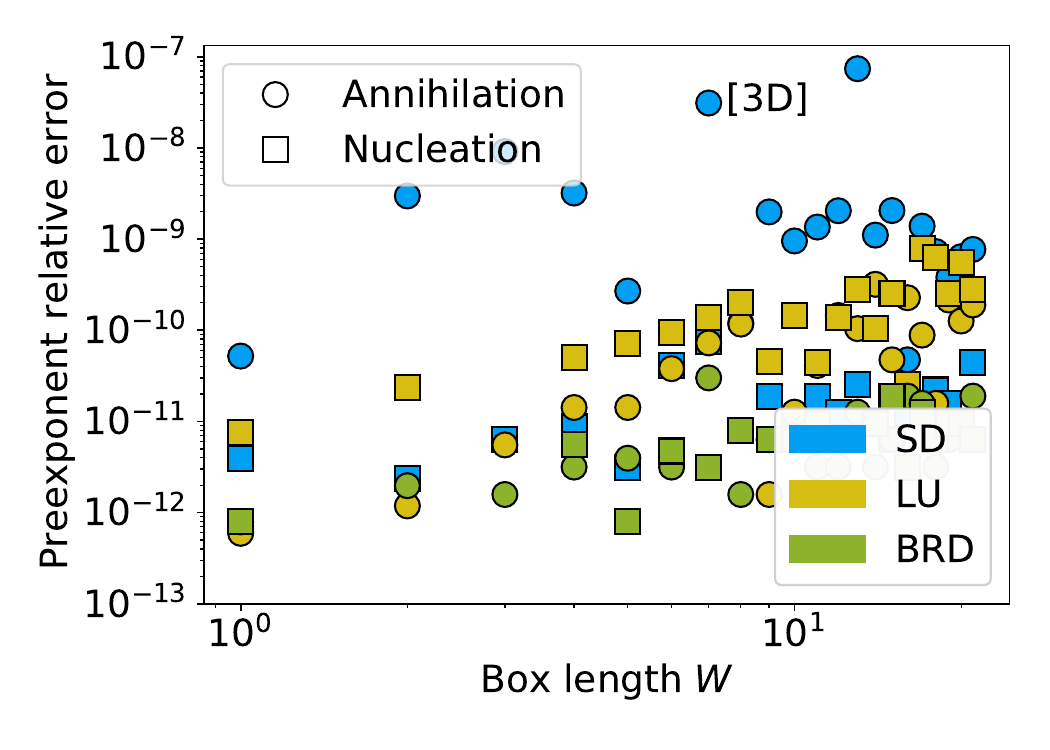"}
  \end{subfigure}
  \caption{Precision of prefactor computation as function of system size.
  Upper and middle panels: two dimensional lattice with free and periodic boundary conditions, respectively.
  Bottom panel: three dimensional system with fixed size in x-y plane and increasing thickness $W$. 
    Box sizes are the same as in Fig. \ref{fig:time}.
  }
  \label{fig:rate}
\end{figure*}

The  lifetime of the magnetic states in HTST derived in \cite {ref6} is expressed in terms of the eigenvalues of Hessian of energy, in particular, the dynamic pre-exponential factor contains the eigenvalues that explicitly requires the computation of spectral decomposition of the Hessian.
Spectral decomposition was used to calculate the lifetime of Sks
 in  \cite {ref6, ref10, ref18, ref19}.
In a recent article \cite {ref28} it was noted that the dynamic term
can be computed as the value of the quadratic form defined by the Hessian on certain vector in spin representation, avoiding spectral decomposition completely, therefore, 
only determinants in the expression for  prefactor remains a challenge.

Langer's theory gives another expression for the lifetime \cite {ref8, ref16}, which does not include spectral decomposition, but only the determinant of the energy Hessian and the only negative eigenvalue of the linearized Landau-Lifshitz-Hilbert equation at the saddle point.
Nevertheless, to calculate the ratio of determinants in (\ref{eq29}), the
spectral decomposition of  Hessian of energy is usually used \cite{ref6, ref10, ref18, ref19}
Note, however, that in principle the determinant ratio can be calculated without decomposition into eigenvalues, for example, using LU or QR decomposition.

In \cite {ref16} it was noticed that the ratio of ordered eigenvalue of Hessian 
 is almost one  for Sk and TS, except for the smallest eigenvalues. Therefore it seems reasonable to save some efforts skipping computation of all eigenvalues and estimate  prefactor by lowest eigenvalues.
However, the usefulness of this method is limited, since the number of eigenvalues needed to obtain a reasonable precision varies from system to system and can be large.
For example, in the benchmarks below
from $4\%$ (for 2D systems) to $50\%$ (for 3D systems)
of all eigenvalues are required to obtain correct order of the entropy prefactor.

For periodic systems, such as UMS in a square or cubic domain with periodic boundary conditions, the Fourier transform can be applied to diagonalize the Hessian matrix, resulting in a quasi-explicit formula for the determinant in terms of the product of $ 2N $ values, where $ N $ is the number of magnetic moments. Below, we will compare the following methods for the determinant computation: 

\begin{description}
  \item[SD] Spectral decomposition for Hessian 
  used to obtain determinant in the following form:
  $$
  \ln \det|H|=\sum_n \ln|\zeta_n|,
  \quad 
  H=U\mathrm{diag}(\zeta_n)U^T
  .$$
  SciPy routing scipy.linalg.eigh was used to compute spectrum.

  \item[LU] LU decomposition for Hessian matrix $H=LU$  
  gives the following expression for the determinant 
  in terms of diagonal entries $L_{nn}$ of the lower-triangular matrix $L$:
  $$\ln \det|H|=\sum_n \ln|L_{nn}|.$$
  SciPy routine scipy.linalg.det was used as an implementation.
  If a meta-stable state is considered, Cholesky decomposition can be used instead.

  \item[FT] Diagonal $2\times2$ blocks $H_p$ of Fourier transformed
  Hessian matrix for periodic system was explicitly formed,
  then scipy.linalg.det was used to compute determinants of $2\times2$ blocks $H_p$, 
  which were combined after that as follows:
  $$\ln\det|H| = \sum_p \ln|\det H_p|,$$ 
  where $p$ runs over all quasimomentum values.
  Intermediate computations in the case involve complex numbers.
  
  \item[BRD] Determinant of block tridiagonal Hessian matrix
  was computed using described in the previous sections method
  in particular eq. (\ref{eq:detper}) and 
  eq. (\ref{eq:detfree}) for periodic
  and free boundary conditions, respectively.
\end{description}

There are several representations of magnetic moments that define magnetic configurations and energy surface. Among them are spherical coordinates \cite {ref9, ref11},
ATLAS method using stereographic projections \cite {RBBK15},
rotation matrices \cite {IDTUJ20} and 3D unit vectors \cite {DP99, VLDHGW14, FKOSEFSSYKSM17}.
In exact arithmetic, the Hessians must coincide exactly in all approaches. Differences arise only in the choice of the local basis near stationary states. In practice, the accuracy of the representations can be different, for example, spherical coordinates suffer from peculiarities near the poles.
In the benchmark, we use 3D unit vectors to represent the spin orientations, which is a reasonable choice for comparison, since there are no singularities, no basis change is required, which leads to an accuracy not worse than in the above articles.
 
For benchmark we consider three systems of increasing size up to million of atoms.
The first two systems are 2D structures on a square lattice with free boundary conditions \textbf{[2F]} and periodic boundary conditions \textbf{[2P]}. 
The third system is a 3D system with a simple cubic structure, periodic boundary conditions in the x-y plane, 
free boundary conditions along the z, and an increasing number of layers in the z direction \textbf{[3D]}.

\begin{table*}\centering
  \begin{tabular}{c | c | l | r | r | l | l}
    Box size $W\times W'$ & $\Delta E_{decay}$ (J) & $\kappa_{decay}$ (Hz) & $\lg \det H^{m}$ & $\lg \det H^{TS}$ 
      & $\kappa^{ent}$ & $\delta k^{ent}$  \\
    \hline\hline
    $30\times31$ & $1.367$ & $2.00\cdot10^{8}$ & $1012.830$ & $1019.888$ & $2.95\cdot10^{-04}$ & $5\cdot10^{-11}$ \\
    $45\times46$ & $2.060$ & $2.48\cdot10^{7}$ & $2160.415$ & $2169.086$ & $4.61\cdot10^{-05}$ & $2\cdot10^{-10}$  \\
    $70\times72$ & $2.766$ & $2.17\cdot10^{6}$ & $5148.701$ & $5159.410$ & $4.42\cdot10^{-06}$ & $6\cdot10^{-10}$ \\
    $100\times103$ & $3.332$ & $2.15\cdot10^{5}$ & $10444.687$ & $10456.967$ & $7.24\cdot10^{-07}$ & $8\cdot10^{-10}$ \\
    $150\times154$ & $3.888$ & $5.20\cdot10^{4}$ & $23348.843$ & $23362.199$ & $2.09\cdot10^{-07}$ & $2\cdot10^{-09}$ \\
    $230\times236$ & $4.415$ & $8.32\cdot10^{3}$ & $54826.139$ & $54840.556$ & $6.18\cdot10^{-08}$ & $>6\cdot10^{-11}$ \\
    $340\times349$ & $4.825$ & $1.53\cdot10^{3}$ & $119886.573$ & $119901.747$ & $2.58\cdot10^{-08}$ & $>3\cdot10^{-11}$ \\
    $500\times513$ & $5.160$ & $5.04\cdot10^{2}$ & $259283.215$ & $259298.839$ & $1.54\cdot10^{-08}$ & $>2\cdot10^{-10}$ \\
    $750\times770$ & $5.450$ & $2.13\cdot10^{2}$ & $584079.328$ & $584095.183$ & $1.18\cdot10^{-08}$ & $>1\cdot10^{-09}$ \\
    $1000\times1027$ & $5.620$ & $1.31\cdot10^{2}$ & $1039039.601$ & $1039055.468$ & $1.16\cdot10^{-08}$ & $>3\cdot10^{-09}$ \\
  \end{tabular}
  
  \caption{Characteristics of isolated Sk decay in 2F benchmark for various lattice sizes 
  corresponding to the same micromagnetic parameters.
  Transition rate is given by Arrhenius law parametrized 
  by activation barrier $\Delta E_{decay}=E^{TS}-E^{m}$,
  and pre-exponential factor $\kappa_{decay}=const\cdot \kappa^{ent}\kappa^{dyn}$,
  where $\kappa^{dyn}$ is computed in HTST approximation (\ref{eq:dyn}).
  Computation of entropy prefactor $\kappa^{ent}$ involves evaluation
  of Hessian of energy determinants at Sk $\det H^m$ and at (TS) $\det H^{TS}$
  computed by BRD method according to (\ref{eq:detfree}).
  Relative error of the entropy prefactor $\delta \kappa^{ent}$ 
  is estimated as half-sum of absolute errors of the log-determinants,
  which in its turn are estimated by evaluation of the determinant 
  for two systems with swapped axes. 
  Jump in precision is explained by change of number of results used for the estimation, 
  since for the smallest five cases results by SD and LU methods are available, 
  in other cases only BRDm results are compared.
  Prefactor is computed for Heisenberg exchange $J=10\,(meV)$
  and magnetic moment $\mu=3\mu_B$.
  Prefactor determines Sk decay rate for extremely small temperatures 
  (tunnelling is not taken into account).
  Results show higher stability of large Sk both due to 
  the energy barrier and the entropy prefactor.
  }
  \label{tb:sizes2}
\end{table*}

System \textbf{[2F]} is a square lattice of the size $W\times W'$
with free boundary conditions.
The sizes $W$ and $W'$ are slightly different, exact sizes are shown 
in Table \ref{tb:sizes2}; the largest system has $10^6$ atoms or $2\cdot 10^6$ degrees of freedom.
The system parameters for the lattice $30\times31$ are chosen as follows:
$$
D_0=0.35\,J,\quad 
K_0=0.16\,J,\quad
\mu H_0=0.02\,J.
$$
The parameters for larger lattices $W\times W'$ are chosen to preserve 
energies and relative size of the Sks:
$$
D=\lambda D_0,\quad
K=\lambda^2 K_0,\quad
\mu H=\lambda^2\mu H_0,\quad
{\rm where} \quad \lambda = 30/W.
$$
The processes of annihilation and generation of Sks are considered.
Three states are taken into account: Sk, UMS and TS, corresponding to decay within the domain.
The Sk state and TS are shown in Fig. \ref{fig:bench2}.

Second benchmark \textbf{[2P]} is the same system as
\textbf{[2F]} but with periodic boundary conditions.
The same Sk annihilation and nucleation processes are considered,
with the same states as in Fig. \ref{fig:bench2}
except that there is no twist of magnetic moments on the boundary (not shown in  Fig. \ref{fig:bench2}).
In the case of [2P], the Hessian matrix is not strictly tridiagonal, but contains corners, and a different formula for BRDm need be used.
The UMS state is periodic, which allows the FT method to be applied.
In a continuous periodic model, all states can be translated preserving
energy, hence, the Hessian of energy has zero modes for nonuniform states.
Continuous zero modes become quasi-zero modes for discrete Hessians.
Quasi-zero modes tend to zero as the system density tends
to zero (in the $W\to\infty$ test),
therefore for sufficiently large systems HTST theory is not 
applicable, and special care should be taken to estimate the value of the mode.
In the benchmark \textbf{[2P]} Sk state has two quasi-zero modes,
therefore we omit Sk state in the benchmark to avoid zero-modes estimation,
which will be described elsewhere.
Transition state however is only fewer times large than lattice constant,
hence TS has no zero-modes due to lattice effects.

The third benchmark \textbf{[3D]} is a thin film represented  by simple cubic lattice of size $30\times31\times W$ 
with increasing number of layers $W=1\ldots20$.
Boundary conditions are free in direction perpendicular to film and periodic in film plane. 
The Dzyaloshinsky-Moriya vectors are along the bonds.
The external field and easy axis anisotropy are  orthogonal to film plane. 
The system parameters does not depend on the film width $N$:
$$
D=0.35\,J,\quad 
K=0.16\,J,\quad
H=0.02\,J.
$$
Sk tube annihilation process is considered, which is a transition to the UMS 
through the TS, corresponding to the tube separation from one boundary 
for moderate thicknesses of the sample, see Fig. \Ref{fig:bench3}. 
For the BRD method, both the first and the second axes were used for recursion.

All metastable states were computed using non-linear conjugate gradient method 
with line search consisting of single step of Newton method.
Transition states were computed using
TMEP method as described in \cite{ref25}.

The BRD method was applied for systems of all sizes.
SD and LU methods were used only for relatively small systems,
where the dense Hessian matrix can be formed in RAM.
The FT method was applied to the UMS state in the 2P benchmark.
Each benchmark was run exclusively on a dedicated machine,
using 24 threads on Xeon E5-2678 v3 processor for \textbf {[3D]} benchmark and 16 threads on an AMD Ryzen 7 2700 processor for the \textbf {[2P]} and \textbf {[2F]} benchmarks.
All computations are performed in double precision arithmetic. The execution times for determinant computations are presented in Fig. \ref{fig:time}.  The execution time is almost the same for different states and follows to the computational complexity asymptotic  for different methods \cite{IK94}:
\begin{equation}\label{eq:theortime}
T_{SD}=O(N^3),\quad 
T_{LU}=O(N^\omega),\quad
T_{FT}=O(N\ln N),\quad 
T_{BRD}=O(N_1N_2^\omega),
\end{equation}
where $N=WW'$ is number of magnetic moments in the system of size $W\times W'$ (as in definition of the system above),
for BRD $W'$ is size of the block,
and $\omega<3$ is complexity of the fast matrix multiplication method in use.

Since the amount of memory required to store a dense Hessian matrix  grows rapidly with increasing system size, SD and LU method were only used for systems with less than $22500$ magnetic moments (requiring $128$ Gb of RAM for Hessian matrix). Huge memory requirement and high computational complexity make  SD and LU methods impractical for computations on large systems, whereas  BRD can be used for structures of orders of magnitude larger.

Since the exact values of the determinants of the Hessians are unknown for magnetic systems, with the exception of the quasianalytic result for the UMS state, we use the invariance of the determinant with respect to the numbering of the axis of the system to measure the accuracy of the methods.
Namely, we calculated each determinant twice, first for the $W\times W'$ lattice (and $ 30 \times 31 \times W $ in [3D]), 
and secondly, for $W' \times W$ lattice (and $ 31 \times 30 \times W $ in [3D]).
The differences in the logarithms of the determinants are shown in Fig. \ref{fig:det}, 
which can be used as an estimate of the relative error in calculation of the determinants. 
We also compared the results of computation of the determinant by different methods. 
The maximum error was always of the same order of magnitude as the error of the worst methods, estimated by comparing the transposed systems, 
therefore, the comparison of different methods is not shown in the figure to facilitate perception.

Computation of the eigenvalues of a symmetric matrix is a well-conditioned task, and the eigenvalues of the Hessian can be found with a relative error that is only several times higher than the machine precision $\varepsilon$. This means that SD and FT tests must calculate determinants with a relative error $c \cdot N \cdot \varepsilon$, where $ N $ is the number of magnetic moments and $ c $ is a moderate constant. The estimate is used as the upper bound for the error in Fig. \ref{fig:det}. It is seen that the estimate gives the correct asymptotic for the SD and FT methods, as well as for the LU and BRD methods. 
The constant $c$ depends on the method and the state. 
The state Sk has the largest error, while TS and UMS states are  calculated with approximately the same accuracy, since TS and UMS differ only by a few tens of nodes. The Fourier transform is predictably both the most accurate and the fastest method.

Spectral decomposition also showed good accuracy, with the exception of the Sk state, which has quasi-zero modes, 
but this is the slowest of the discussed methods. 
The BRD method showed comparable result to SD method on TS and UMS states,
and an order of magnitude more accurate result than SD for Sk state.
Overall BRDm is among methods demonstrating smallest error, being much faster than LU and SD decomposition. 
The LU method shows the worst accuracy, but more than $10$ times faster than SDm; 
hence LU is a reasonable alternative to BRDm for dense Hessian matrices when BRDm is not applicable.

The transition rate and the lifetime of magnetic systems contain the entropy prefactor $\kappa^{ent}$, which is the ratio of the determinants of the Hessians. 
The value of the logarithm of the determinant increases in proportion to the number of spins and is huge for large systems
as shown in table \ref{tb:sizes2}. 
At the same time, the logarithm of the lifetime is weakly dependent on the density of the lattice under proper scaling. 
Therefore, the problem of calculating the entropy prefactor for large systems is a difficult task, 
since for large systems the determinants must be calculated with higher accuracy. 
Taking into account the accuracy of calculating the determinants, as estimated above, 
we show the error for calculating the entropy prefactor using different methods in Fig. \ref{fig:rate}. 
The BRD method has shown better accuracy than SD and LU methods, as well as better performance in terms of speed.

\section{Conclusions}

Exploiting the sparse  structure of the Hessian matrix for energy in the Heisenberg-like model with short-range interaction, we developed the BRD method for fast computing the determinant of the Hessians, and have applied the results to calculate the lifetimes of magnetic systems in HTST.

The method has relatively small memory requirements, and
at any  time, only a block corresponding to one layer of the system should be stored in memory. The computational complexity of the method is linear in the number of layers, which makes it possible to use the method for calculating the lifetimes of large systems with millions of magnetic moments systems with millions of spins. 
In some cases the BRD method can be used for computation of systems with billions of atoms, specifically for elongated systems: for a $10^9\times 10$ lattice, 
computation time is of the same order as for $10^3\times 10^3$.

Above, we gave explicit formulas for computation of the determinants of block tridiagonal matrices, which can be easily generalized to a general banded matrix. 
The formulas are recursive, and we computed all the required matrices layer by layer, using parallelism only for block multiplication and for matrix inversion.
Further improvements of the performance can be achieved implementing parallel techniques,
e.g. see \cite{BCR88, MLP94, KN14, N18}.
  
In the benchmarks above, we excluded from the analysis Sk structures with periodic boundary conditions, due to the existence of zero modes, that is, state transformations with conservation of its energy.
HTST is inapplicable in this case; however, in the further development of the theory, one should take into account the zero
or quasi-zero modes, when the eigenvalues of the Hessian matrix are close to zero.

In this case, the integrals in the section \ref{sec:rate}
cannot be estimated for infinite regions, but the volume of zero modes should appear in the answer. Moreover, if the number of zero modes is different in the minimum and TS, then the prefactor in the Arrhenius law will depend on the temperature.
Sometimes the volumes of the zero modes can be easily estimated, as for the translational modes. However, in general (for example, rotation modes for a pair of Sks,
oscillatory modes for a Sk tube, or quasi-zero modes for Hopfions) this is a complex problem beyond the scope of this article. 
This problem will be addressed elsewhere.

\section*{Acknowledgements}
We thank Maria Potkina for helpful discussion.
This work was funded by Russian Science Foundation (Grant 19-42-06302).

\bibliography{ms}

\end{document}